\documentclass[sn-basic]{sn-jnl}

\usepackage{graphicx}%
\usepackage{multirow}%
\usepackage{amsmath,amssymb,amsfonts}%
\usepackage{amsthm}%
\usepackage{mathrsfs}%
\usepackage[title]{appendix}%
\usepackage{xcolor}%
\usepackage{textcomp}%
\usepackage{manyfoot}%
\usepackage{booktabs}%
\usepackage{algorithm}%
\usepackage{algorithmicx}%
\usepackage{algpseudocode}%
\usepackage{listings}%
\usepackage{sidecap}[raggedright]

\begin{document}

\title{GeoCoDA: Recognizing and Validating Structural Processes in Geochemical Data. A Workflow on Compositional Data Analysis in Lithogeochemistry}

\author*[1]{\fnm{Eric} \sur{Grunsky}} \email{egrunsky@gmail.com}%

\author[2]{\fnm{Michael} \sur{Greenacre}} %\email{michael.greenacre@upf.edu}%

\author[3]{\fnm{Bruce} \sur{Kjarsgaard}} %\email{bakjaarsgard@gmail.com}

%\thankstext{t1}{Some comment}

\affil*[1]{\orgdiv{ Department of Earth and Environmental Sciences}, \orgname{University of Waterloo},
 \orgaddress{\city{Waterloo}, \country{Canada}}}
\affil[2]{\orgdiv{Department of Economics and Business, and Barcelona School of Management}, \orgname{Universitat Pompeu Fabra},
\orgaddress{\city{Barcelona}, \country{Spain}}}
\affil[3]{\orgname{Geological Survey of Canada, Natural Resources Canada}, \orgaddress{\city{Ottawa}, \country{Canada}}}

\abstract{
Geochemical data are compositional in nature and are subject to the problems typically associated with data that are restricted to the real non-negative number space with constant-sum constraint, that is, the simplex. Geochemistry can be considered a proxy for mineralogy, comprised of atomically ordered structures that define the placement and abundance of elements in the mineral lattice structure. Based on the innovative contributions of John Aitchison, who introduced the logratio transformation into compositional data analysis, this contribution provides a systematic workflow for assessing geochemical data in a simple and efficient way, such that significant geochemical (mineralogical) processes can be recognized and validated. This workflow, called GeoCoDA and presented here in the form of a tutorial, enables the recognition of processes from which models can be constructed based on the associations of elements that reflect mineralogy. Both the original compositional values and their transformation to logratios are considered. These models can reflect rock-forming processes,  metamorphism, alteration and ore mineralization. Moreover, machine learning methods, both unsupervised and supervised, applied to an optimized set of subcompositions of the data, provide a systematic, accurate, efficient and defensible approach to geochemical data analysis. The workflow is illustrated on lithogeochemical data from exploration of the Star kimberlite, consisting of a series of eruptions with five recognized phases.}


\maketitle

%%%%%%%%%%%%%%%%%%%%%%%%%%%%%%%%%%%%%%%%%%%%%%%%%%%%%%%%%%%%%%%%%%
\section{Introduction}

One of the primary purposes of geochemical data analysis is the recognition of geochemical/geological processes. Processes are recognized by a continuum of variable responses and the relative increase/decrease of these variables.  The compositional nature of geochemical data requires careful consideration when examining relationships between the elemental, oxide or molecular constituents that define a composition. In materials, rocks or derived from bedrock, these constituents are proxies for mineralogy. 

The use of ratios are essential when making comparisons between elements in processes such as igneous fractionation \citep{Pearce:68,Stanley:19, Stanley:20, Urqueta:1}. Furthermore, the logarithms of ratios, called logratios, are essential when measuring moments such as variance/covariance in the examination of data derived from geochemical surveys \citep{Aitchison:86, Buccianti:06, PawlowskyBuccianti:11, BoogaartTolosana:13, Greenacre:18, Greenacre:21}.
The use of ratios also satisfies the principle of subcompositional coherence, which means that the results are not particular to the choice of elements included in the study \citep{Aitchison:86, Grunsky:30}.

\cite{Aitchison:99} stated that the relationships between the elements of geochemical data are governed by ``natural laws”, which implies that these relationships are governed by stoichiometry. 
\cite{GrunskyBaconShone:10} have shown that geochemical patterns and trends are closely related to the stoichiometric constraints of minerals.

To effectively interpret geochemical data, a two-phase approach is suggested; that of process discovery, followed by process validation \citep{Grunsky:32}. This tactic identifies geochemical/geological processes that exist in the data, which are not obvious unless appropriate transformations and/or statistical methods are utilized. The process discovery phase is most effective when carried out using a multivariate approach. Linear combinations of elements related by stoichiometry are generally expressed as strong patterns, whilst random patterns and under-sampled processes show weak or uninterpretable patterns. \cite{Grunsky:29} initially demonstrated these concepts using multi-element lake sediment geochemical data from the Melville Peninsula area, Nunavut, Canada.

A straightforward example of assessing lithogeochemical data is given in this manuscript. 
We demonstrate the usefulness of compositional data analysis, using univariate, bivariate and multivariate methods, for process discovery and validation from the Star kimberlite in Saskatchewan, Canada. Distinct geochemical kimberlite phases can be statistically identified using this approach and lead to efficiencies in the economic evaluation of kimberlite for diamonds \citep{Grunsky:30}. Subsequent statistical applications of kimberlite whole rock geochemistry in relation to diamonds have been undertaken at the Orapa diamond mine by \cite{Stiefenhofer:1} and at the Attawapiskat kimberlite field-Victor diamond mine by \cite{Januszczak:1}. 

%%%%%%%%%%%%%%%%%%%%%%%%%%%%%%%%%%%%%%%%%%%%%%%%%%%%%%%%%%%%%%%%%%
\section{Materials and methods}

\subsection{The kimberlite data}
%\cite{Grunsky:30} evaluated diamond drill core lithogeochemistry from the Star kimberlite located in Saskatchewan, Canada.
The Star kimberlite is a series of  distinct kimberlite eruptions with five recognized phases (Cantuar, Pense, early Joli Fou (eJF), mid Joli Fou (mJF), and late Joli Fou (lJF)). 
%Grunsky and Kjarsgaard (2008) evaluated drill core lithogeochemistry from the Star kimberlite located in Saskatchewan, Canada. The Star kimberlite is a series of kimberlite eruptions with five recognized phases (Cantuar, Pense, early Joli Fou (eJF), mid-Joli, Fou (mJF), and late-Joli Fou (lJF). 
The early Joli Fou phase was originally recognized to contain more macro-diamonds than the other phases, thus making it useful to recognize this phase in the diamond exploration and evaluation process. Subsequent drilling, large diameter drilling and underground bulk sampling \citep{Harvey:1, NI43101:1} subdivided the Cantuar kimberlites into north and south sub-phases and also identified a (non-resource) late stage juvenile lapilli-rich pyroclastic kimberlite (JLRPK).   Revised diamond grades for the Cantuar-S and eJF are similar at 18ct/100t, followed by the Pense (13 ct/100t) and mJF (7 ct/100t), however, the diamond sample parcel value (US\$/ct) of the Cantuar-S is approximately double that of the eJF (NI 43-101, 2018).  Hence it is highly desirable to recognize the early Joli Fou and Cantuar kimberlite phases in the diamond exploration and evaluation process.

The data set considered in the present work consists of geochemical analyses of 270 of these kimberlite samples, with 22 elements identified and classified as major, minor, trace or rare:
\begin{itemize}
    \item major: Si, Mg, Fe, Ca, Al, Ti, Na, K, P
    \item minor: Ni, Cr, Co, Rb, V
    \item trace: Ga, Th, Nb, Zr 
    \item rare: La, Y, Er, Yb 
\end{itemize}

This data set has no zero values, or values reported below the lower limits of instrument detection, which is not a common situation in compositional data analysis in most fields of study, but simplifies our present task when we deal with element ratios and logarithmic transformations.
The issue of compositional zeros will not be dealt with in detail in this study, but is discussed in the final section.
Another aspect of these data that makes their analysis easier is that we are not considering the geospatial information (easting, northing, depth) of the samples in this study. 
The analysis of geochemical compositions that are spatially related will be the subject of a follow-up article tackling a more complex data set. 

Our workflow for analyzing such a lithogeochemical compositional data set is called GeoCoDA (= geochemical compositional data analysis).
The only difference between analysing compositional data and the usual statistical approach to analyzing a multivariate data set, is the question of the transformations performed on the compositional data. The statistical methods used are mostly standard and well-known, but then care has to be taken in the interpretation of the results in the light of the data transformations that were made initially.

\subsection{Process discovery}
Process discovery, or process identification, involves the use of both univariate, bivariate and multivariate statistical methods. Process discovery with geochemical data reflects the recognition of linear models that reflect the stoichiometry of rock-forming minerals and subsequent processes that modify mineral structures (e.g., hydrothermal fluids, metamorphism, weathering, groundwater interaction). Additional processes such as surface water fluid dynamics effects can effectively sort minerals according to the energy of the environment and mineral density. The chemistry of minerals is governed by stoichiometry and the relationships are easily described within the simplex. In the line of Aitchison's approach, geoscientists have long recognized that many geochemical processes can be clearly described using cation ratios that reflect the stoichiometric balances of minerals during formation (e.g. \cite{Pearce:68}).
In the language of machine learning, the methods of process discovery can be termed \emph{unsupervised learning}.

%Geochemical data, expressed in elemental form
As geochemical data are a proxy for mineralogy, if the mineralogy of a geochemical data set is known, then the proportions of these elements can be determined using linear methods. \cite{Grunsky:33} reviewed some of the normative mineral calculation procedures that are available. Normative mineral calculations can be complicated due to the continuum of element substitutions (e.g.,  simple substitutions such as Ca:Na +  K in plagioclase, or Fe:Mg in olivine, or more complex substitutions such as Si:Na + Al in pyroxenes). 
This requires assumptions to be made about such compositions, and the resulting estimates may not reflect the actual mineral compositions or abundances. Thus, in this study, we have chosen to use only the geochemical data and use the observed linear patterns as proxies for mineralogy.

An essential part of the process discovery phase is to recognize the problem of closure  (that is, expressing the data as proportions), and to choose a suitable logratio transformation -- see Section 3.1 for a practical explanation of the available transformations and Supplementary Material, Section S1, for a summary of their theoretical definitions.

%%%%%%%%%%%%%%%%%%%%%%%%%%%%%%%
\subsection{Process validation}
%Process validation is the methodology employed to verify that a geochemical composition (response) is associated with identified processes. The processes can be in the form of e.g., lithology, soil character, ecosystem properties, climate, or deeply buried tectonic assemblages \citep{Gallagher:1, Grunsky:31}. Validation can be in the form of an estimate of likelihood that a composition can be assigned membership to one of the identified processes. This is typically done through the assignment of class identifier or a measure of probability. Assignment of class membership can be done through the application of techniques such as discriminant analysis, logistic regression, neural networks, classification trees or random forests \citep{Hastie:09}. In the language of machine learning, process validation can be termed \emph{supervised learning}.

%An essential part of process validation is the selection of variables that enable efficient classification, which involves selecting variables that maximize the differences between the different classes and minimizes the amount of overlap due to noise or unresolved processes in the data. Within the context of compositional data, variables that are selected for classification require transformation to logratio coordinates. 
%Here, the additive logratio transformation can be useful, since it relies on a common denominator and is thus easier to interpret \citep{GreenacreMartinezBlasco:21, GreenacreEtAl:23}.

Process validation is the methodology employed to verify that geochemical compositions are associated with identified processes. The processes can be in the form of, for example, lithology, soil character, ecosystem properties, climate, or deeply buried tectonic assemblages \citep{Gallagher:1, Grunsky:31}. Validation can be in the form of an estimate of likelihood that a composition can be assigned membership to one of the identified processes through the assignment of class identifier or a measure of probability.
Assignment of class membership can be done through the application of techniques such as discriminant analysis, logistic regression, neural networks, classification trees or random forests \citep{Hastie:09}. In the language of machine learning, process validation can be termed \textit{supervised learning}. 

An essential part of process validation is the selection of variables that enable efficient classification, which involves selecting variables that maximize the differences between the different classes and minimizes the amount of overlap due to noise or unresolved processes in the data. A transformation to logratio coordinates is necessary to classify compositional data.

Classification results can be conveniently expressed as direct class assignment or the posterior probabilities in the form of forced class allocation, or as class typicality. 
The suitability of an assignment to a given class can be assessed by examining the quantitative measure of assignment such as posterior probability. If the measure of assignment is below some arbitrarily defined amount, then the classification can be ignored.
Classification accuracies can be assessed through the generation of tables or graphical displays (e.g., a ROC curve; \cite{Hastie:09}) that show the accuracy and errors measured from the estimated classes against the initial classes in the training sets used for the classification. 
%An alternative single numerical measure of classification accuracy is Cramer's V measure of association \citep{Sheskin:97}, applied to the $G\times G$ cross-table of predicted and actual class assignments, where $G$ is the number of groups. 
%Cramer's V is the square root of the chi-square statistic divided by the sample size $N$, called the inertia in correspondence analysis, and further divided by $G-1$: $\chi^2 / \left( N (G-1) \right)$. This measure lies between 0 (no association, assignment purely random) and 1 (perfect association, 100\% accurate assignment).

%%%%%%%%%%%%%%%%%%%%%%%%%%%%%%%%%%%%%%%%%%%%%%%%%%%%%%
\section{GeoCoDA: a workflow for lithogeochemical data analysis}
\begin{figure*}[t]
\begin{center}
\includegraphics[width=13cm]{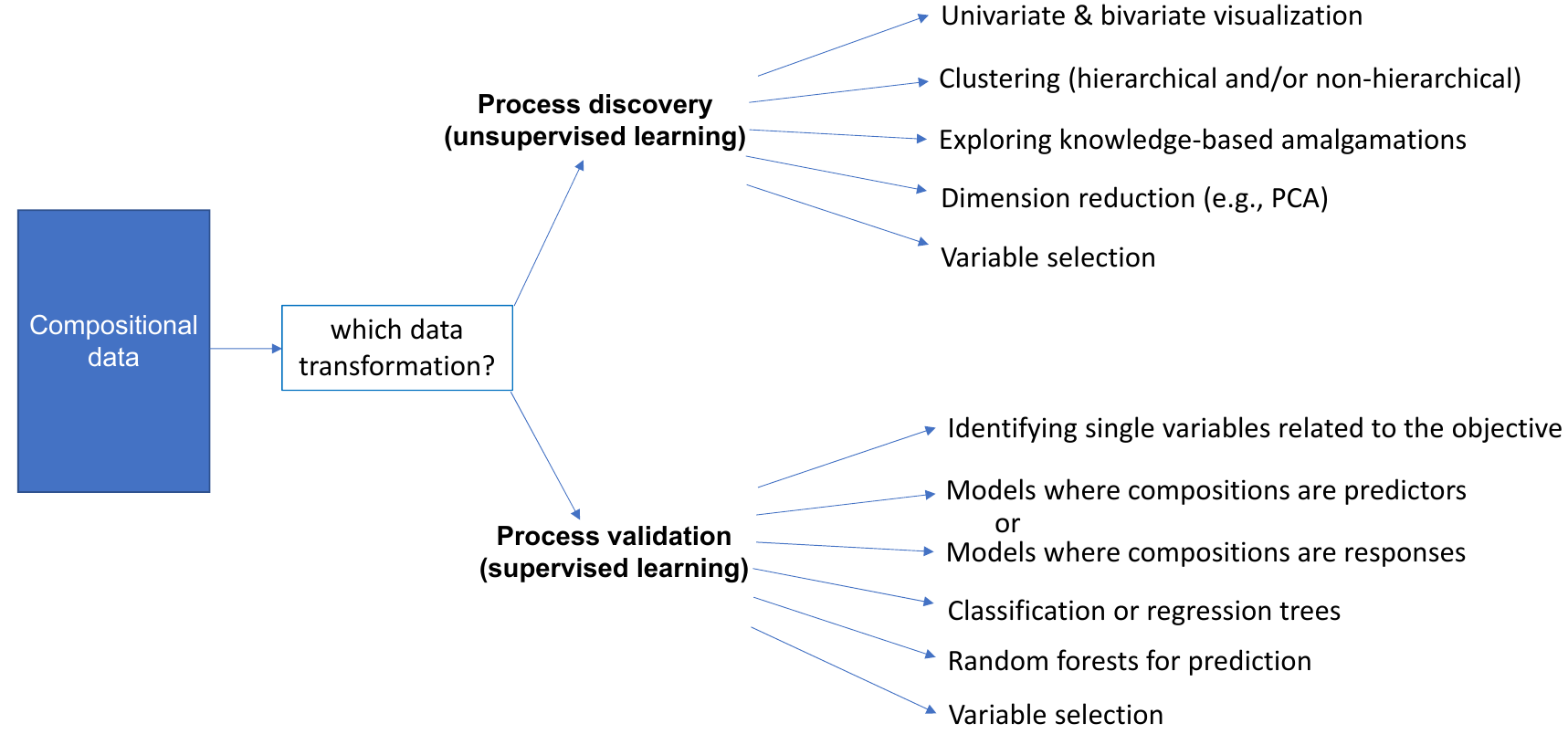}
\caption{GeoCoDA workflow for the analysis of a lithogeochemcial data set. The most important step is the choice of data transformation, after which regular statistical analyses, univariate, bivariate and multivariate are applied. Attention is then paid to the interpretation of the transformed variables in the results.}
\label{WorkFlow}
\end{center}
\end{figure*}

In this workflow, applied to the kimberlite data, we make a clear distinction between process identification and process validation, which in statistical terms translates to unsupervised and supervised learning, respectively (Figure \ref{WorkFlow}). 
In unsupervised mode we aim to understand the data structure \emph{per se}, identifying which are the important elements and element ratios explaining this structure. 
In supervised mode we aim to understand which part of the data structure is related to ``external" variables, which in this application is exemplified by the five kimberlite phases.
In other geochemical applications, additional variables such as spatial position and environmental conditions can explain, or be explained by the geochemistry. External categorical variables are considered as attributes that are associated with the geochemical composition, but are not part of it.

%%%%%%%%%%%%%%%%%%

An important aspect of the dataset is that the major element data, in weight \% oxide, have been converted to elemental (cation) form (oxygen-free), in parts per million (ppm). Thus, all the geochemical data, regardless of being major, minor, trace or rare elements, are expressed in the same units \citep{Grunsky:30}.
Throughout this workflow, the importance of visualization of both the data and the results of the analyses is highlighted.

\begin{figure*}[ht]
\begin{center}
\includegraphics[width=13cm]{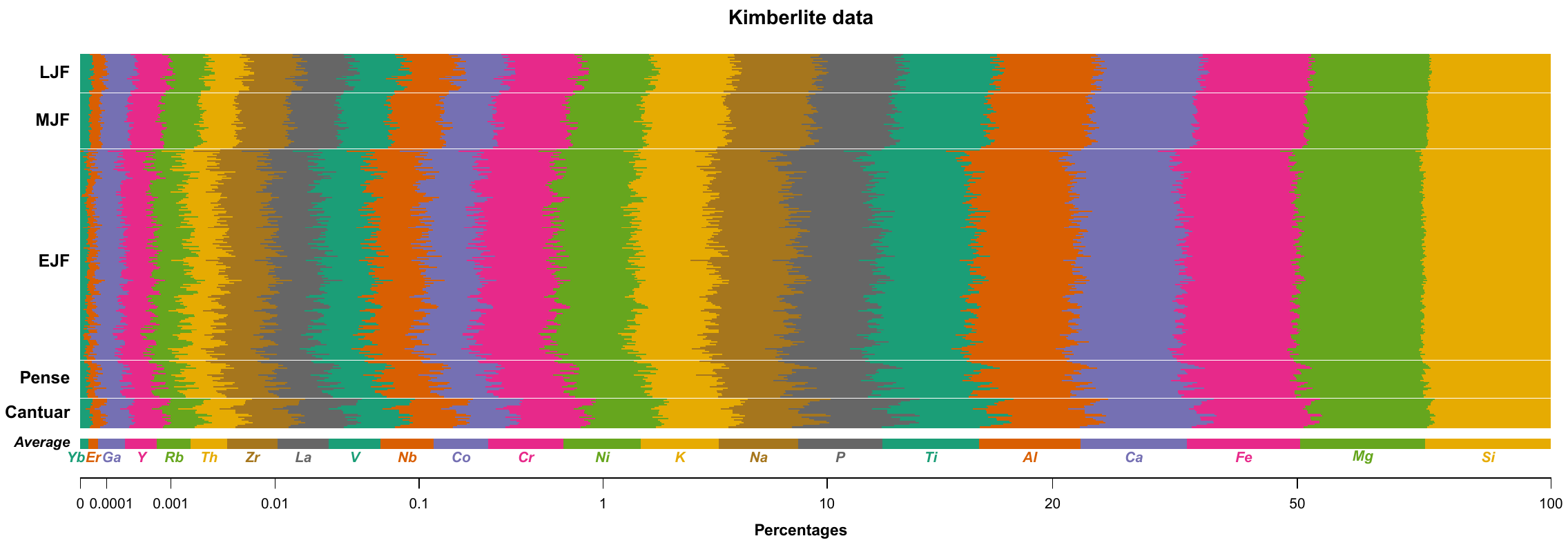}
\caption{An attempt to show all the compositional data, in the form of a compositional bar plot, where elements have been ordered in increasing mean value. The horizontal scale has been monotonically transformed in a nonlinear way, to stretch out the values of rare (i.e. low concentration) elements at the lower end (left) of the scale so that they can be compared in the same way as abundant elements in the upper part (right). The average composition across the whole data set is shown as a separate bar at the bottom. The five kimberlite phases have been ordered from oldest (Cantuar) in the lower rows to youngest (lJF) at the top. 
The nonlinear transformation was achieved by (i) multiplying the compositional data by the inverse of the smallest value in the whole data set, which makes the smallest value 1 while preserving all the relative values (i.e., ratios) in the data, then (ii) performing a logarithmic transformation to pull down all the high values, and finally (iii) closing the data set.}
\label{CompPlot}
\end{center}
\end{figure*}

Before proceeding to in-depth analysis we can think of how to visualize the raw data, while bearing in mind the issue of the inherent lack of subcompositional coherence in such data with the unit-sum property. 
The compositional bar plot in Figure \ref{CompPlot} is an attempt to show all the data, both the very rare and very abundant elements. 
The percentage scale has been transformed nonlinearly, but still monotonically, to exaggerate the scale of the rare elements at the left, so that they can be compared across the phases just like the abundant elements at the upper end of the scale to the right (the elements are ordered from left to right in increasing mean values).
The phases have been grouped and in time order (from top, youngest, to bottom, oldest) and some differences can be clearly seen. For example, the bars of the eJF phase samples all appear pushed to the left by the higher values of the abundant elements Ca, Fe, Mg and Si. This means that ratios with these abundant elements in the numerator and rarer elements in the denominator will tend to have higher values in the eJF samples -- in fact, we will see later that Mg/Zr or Mg/La, for example, are important ratios and are high in the eJF phase.
Identification of such ratios is an important aspect of both process discovery and validation and will be conducted more formally in the next section.
In Figure S1 of the Supplementary Material the compositional bar plot without scale transformation is shown, where only differences in the top abundant elements can be distinguished and compared. 

%%%%%%%%%%%%%%%%%%%

%%%%%%%%%%%%%%%%%%%%%%%%%%%%%%%%%%%%%

\subsection{Logratio transformations} 
A composition is a set of nonnegative numbers that sum to 1.
Thus, zero values are included in a composition but for purposes of computing logratio transformations, all values have to be positive, so zeros need to be replaced (see the Discussion in Section 4).

The simplest logratio transformation of a composition is the pairwise logratio (PLR), the logarithm of the ratio between two compositional parts, log(X/Y);
for example, in the present application between two elements, log(Mg/Zr).
Since there are $J=22$ elements in this data set, there are $\frac{1}{2} \times J \times (J-1) = \frac{1}{2} \times 22 \times 21 = 231$ distinct PLRs.
Notice that there are $J \times (J-1) = 22 \times 21 = 462$ possible logratios but half of them are just the negatives of the others, e.g. log(Mg/Zr)$ = -$log(Zr/Mg).  
Other logratio transformations make specific choices of the denominator of the ratio. 

In order to limit the number of PLRs to consider, the additive logratio (ALR) transformation is the set of PLRs using a fixed chosen denominator, called the reference; for example, the PLRs with Si as the reference, log(X/Si), where X$ \neq $ Si. Thus, there are $J-1 = 21$ PLRs in a particular ALR set, but there are $J = 22$ possible choices for the reference and thus 22 possible ALR sets. 
The choice of the reference in the ALR is based on both statistical and geochemical considerations, since this choice can highlight specific relationships/trends in the data that otherwise might not be apparent. 
In addition, since log(X/Y) $ = $ log(X)$ - $ log(Y), if the logarithm of the reference part, log(Y), has very low variance that is it is almost constant across the data set, then the interpretation of the ALR is, for all practical purposes, the interpretation of log(X) shifted by a constant approximately equal to minus the mean of log(Y). 

Finally, to limit the logratios to a single set of possibilities, the centered logratio (CLR) transformation is a set of $J$ logratios where each is the logarithm of the compositional parts in turn with the geometric mean of all of them serving as reference. 
The CLRs are thus similar to ALRs log(X/Y), but all the parts are used with X as numerators and the reference Y is the geometric mean of all the parts, not a specific selected part.
Hence, there is only one set of $J$ CLRs, and this transformation has many convenient properties: for example, the single set of $J$ CLRs is equivalent in many cases to analysing all $\frac{1}{2} \times J \times (J-1)$ PLRs.
As for ALRs, if the geometric means of all parts across the data set are nearly constant, the set of CLRs can be approximately interpreted in terms of the logarithms of the parts themselves.

An equivalent definition and interpretation of a particular CLR with numerator part X, is that it is the average of all the $J$ PLRs (1/$J$)[ log(X/Y$_1$) + log(X/Y$_2$) + $\ldots$ + log(X/Y$_J$) ], where Y runs over the set of all parts, including X itself. 
Since log(X/X) = 0, the CLR is a small scale factor of $(J-1)/J$ different from the average of the PLRs log(X/Y) where Y runs over $J-1$ parts, excluding X. 
Notice that this average is just the negative of the average of the $(J-1)$ ALRs log(Y/X) where X is the reference part, so the CLRs and ALRs are intimately related. 

We recommend CLRs or simple PLRs, of which ALRs are a special case, as sufficient for statistical analysis \citep{GreenacreEtAl:23}.
We also allow amalgamations of elements (simple sums of compositional values) to be used in ratios with other elements or with other amalgamations, if such ratios make geochemical sense -- see Section 3.3.  
All logratios need special care in their interpretation, as we will demonstrate in the examples to come.

\subsection{Univariate and bivariate analysis}
\begin{figure*}[b]
\begin{center}
\includegraphics[width=13cm]{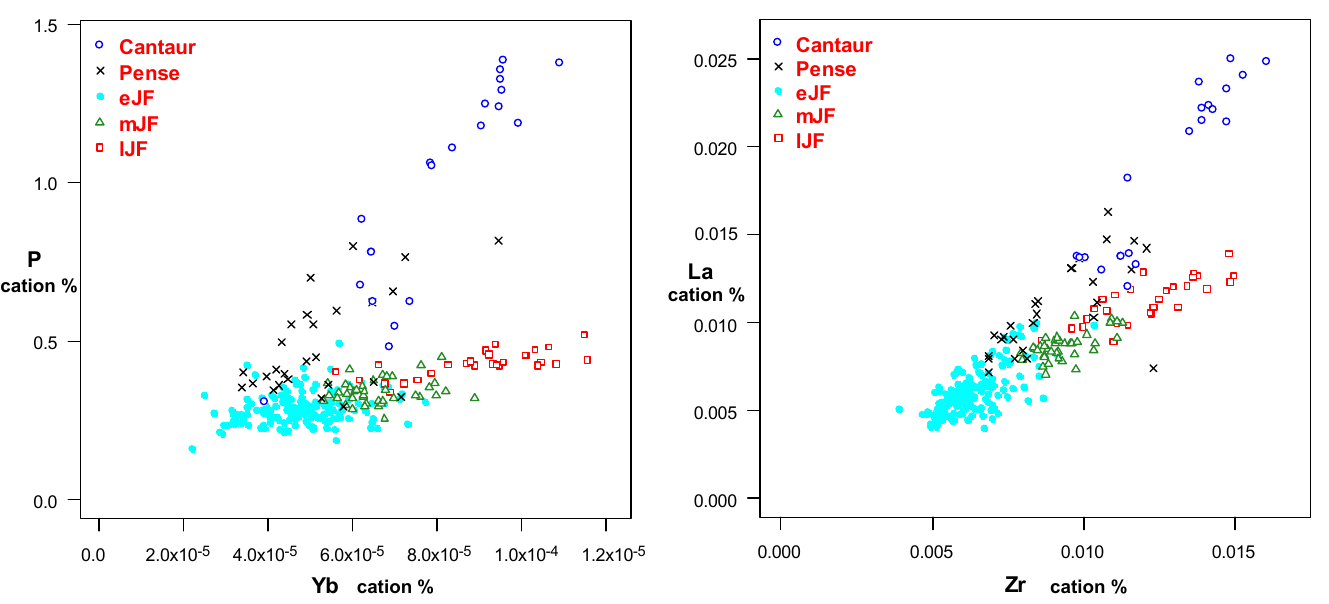}
\caption{Plots of two pairs of elements that are typically associated with kimberlite magma and affected by fractionation (and to a lesser degree) contamination processes from the Star kimberlite geochemistry. Note the linear relationships between these elements within different phases. There is one linear trend associated with the evolution of the earlier Cantuar and Pense eruptions and a second linear trend associated with the eJF-mJF-lJF eruptions.}
\label{ScatterPlot}
\end{center}
\end{figure*}
Compositional data are essentially multivariate, but the initial use of univariate and bivariate approaches can already indicate important aspects of the data set.
Before we pass to logratios, let us see what information we can get from the raw compositions.
Figure \ref{ScatterPlot} shows scatterplots of two pairs of elements that are typically associated with kimberlite magma fractionation and contamination processes. 
The data are presented in cation percent and are untransformed. These plots reveal two distinct patterns. There is a linear pattern of relative enrichment/depletion for all five phases. Additionally, there is a distinct pattern of relative enrichment/depletion associated with the early-, mid- and late-Joli Fou kimberlites. A separate trend is noted for the Pense and Cantuar phases. The greatest relative enrichment of the four elements in Figure \ref{ScatterPlot} is shown by the late Joli Fou and the Cantuar kimberlites. These linear patterns reveal the control that stoichiometry has over the formation of minerals and the associated changes in mineralogy during kimberlite magma contamination (mantle and crustal), and fractionation processes, as described below. Such patterns are well documented by \cite{Pearce:68} and are called Pearce element ratios, abbreviated as PER.

Such linear patterns are also known as evidencing \textit{proportionality} between elements, often used as a measure of association \citep{Lovell:15, Quinn:17}.
Two elements that are perfectly proportional have constant ratios, so can be identified by looking at the variance of their ratios within each phase, or perhaps better, the variance of their logratios, since the log-transformed values are interval-scale.
%Since interest is in linearity within phases, these variances are computed for each phase, and then low variances in some or all phases are sought. 

Coming now to pairwise logratios (PLRs), we have already seen in Section 3.1 that $J(J-1)/2 = 231$ unique PLRs of the $J=22$ elements can be computed. 
To get an idea of their usefulness and importance in both unsupervised and supervised analyses, they can be ordered in several ways, depending on the objective.
Since logratio-transformed compositional data are essentially multivariate, the notion of their total variance is a fundamental quantity, which is either broken down into part contributions due to the PLRs, or the PLRs can be quantified by how much variance they explain, both in the data set itself or in external response variables.

The total logratio variance is equal to 0.1132 in this example, and every logratio makes a separate contribution to this total.
The \textit{contributed variances} thus have a sum equal to 0.1132.
%Seeing which logratios contribute more or less to this total will indicate which logratios are the ones that are more likely to count in any future multivariate analysis. 
This type of computation can be done for the CLRs as well, decomposing the same total variance into contributions from each CLR.
%These CLR contributions would indicate which elements are likely to count more or less in a multivariate analysis (see Section S1 of the Supplementary material for logratio definitions).
Since the idea of principal component analysis (PCA), for example, is to explain as much of the total variance as possible, it is the logratios (PLRs or CLRs)  with higher variance contributions that the analysis would tend to focus more on (for a recent review of PCA, see \cite{GreenacreEtAl:22}).

An alternative ordering of the logratios is in terms of \textit{explained variances}, which are $R^2$ values in the regression sense.
Each logratio explains not only its own contribution to variance but many parts of the variance of all the other logratios, since the logratios are all inter-correlated, often very strongly. 
Hence, all the logratios can be regressed on a specific logratio, and in each case the variances explained are collected and summed to give the part of the total logratio variance that is explained by that particular logratio.
Then the logratios can be ordered from those that explain the most to the least.
Hence, each logratio's explained variance $R^2$ is based on a separate analysis, and they clearly do not sum to 100\%, as do the percentage values of the contributed variances.

Then, a third way of sorting the logratios, similar to the second way above but more focused on a specific objective, is to restrict attention to some specific part of the data variance to be explained.  
In this example it would naturally be the variance between the five phases that is to be explained -- this variance between groups is less than the total variance of the individual samples.
In the kimberlite data, the between-group variance (i.e., between-phase variance) is in fact 40.2\% of the total variance in this data set, with the other 59.8\% being the variance within the five phase groups, which is unrelated to the differences between the groups. 
Again, each logratio (or CLR, or whatever logratio transform is of interest) explains different percentages of this between-group variance, just as in analysis of variance.
This third way is related to a supervised learning objective, identifying logratios that give the best explanation of phase group differences, whereas the first two ways described above are related to the unsupervised learning on the sample differences. 

The top 10 logratios in terms of (i) contributed variance, (ii) explained variance and (iii) between-phase variance are listed in the columns of Table \ref{Ratios}, in descending order of importance.

\smallskip
\begin{center}
\begin{table*}[ht]
\hskip0.5cm\begin{tabular}{lclclc}
\multicolumn{2}{c}{\em (i) }\qquad & \multicolumn{2}{c}{\em (ii)  }\qquad  & \multicolumn{2}{c}{\em (iii)}\\
\multicolumn{2}{c}{\em Contributed}\qquad & \multicolumn{2}{c}{\em  Explained }\qquad  & \multicolumn{2}{c}{\em Explained}\\
\multicolumn{2}{c}{\em variances}\qquad  & \multicolumn{2}{c}{\em variances}\qquad  & \multicolumn{2}{c}{\em variances}\\
\multicolumn{2}{c}{\em (decomposition)}\qquad  & \multicolumn{2}{c}{\em between samples}\qquad  & \multicolumn{2}{c}{\em between phases}\\
 Ratio \qquad & \qquad \% \qquad \qquad  & Ratio  \qquad & \qquad \% \qquad \qquad & Ratio \qquad &  \% \\[2pt]
\hline 
%\rule{0pt}{-0.5ex}\\
Na/Ni &  1.78  & Mg/K  &   47.28 & K/Co\qquad\qquad  &   56.44 \\ 
Mg/Na &  1.76  & Fe/K  &   47.11 & Fe/K   &   56.16 \\ 
Na/Co &  1.74  & K/Co  &   46.77 & K/Ni   &   56.00 \\
Ca/Na &  1.69  & K/V   &   46.53 & V/Yb   &   55.91 \\
Na/P  &  1.66  & K/Cr  &   46.00 & Mg/K   &   55.67 \\
Na/La &  1.66  & Rb/V  &   45.97 & Rb/Co  &   55.62 \\
Na/Nb &  1.61  & Ti/K  &   45.66 & Cr/Ga  &   55.56 \\
Fa/Na &  1.61  & Mg/Rb &   45.55 & Al/Mg  &   55.44 \\
Ti/Na &  1.56  & Fe/Rb &   45.52 & Fe/Rb  &   55.37 \\
Na/Th &  1.51  & Si/K  &   45.41 & Rb/Ni  &   55.03 \\
%Na/V  &  1.50  & K/Ni  &   83.19 & Nb/Co  &   76.00 \\
%Si/Na &  1.48  & K/Nb  &   82.33 & Co/La  &   75.64 \\
%Na/Cr &  1.46  & Rb/Co &   82.01 & Ni/Y   &   75.62 \\
%K/Ni  &  1.34  & Ti/Rb &   81.72 & Th/Ni  &   75.28 \\
%Na/Zr &  1.34  & Rb/Nb &   81.12 & V/Ni   &   75.20 \\
%K/Co  &  1.29  & K/P   &   80.79 & Zr/Co  &   75.02 \\
%K/Nb  &  1.27  & Ti/Na &   80.75 & Fe/La  &   74.74 \\
%K/La  &  1.27  & Rb/Cr &   80.69 & Si/Y   &   74.35 \\
%Mg/K  &  1.26  & K/La  &   80.18 & Fe/Nb  &   74.25 \\
%Na/Y  &  1.25  & Na/Nb &   80.13 & Fe/Zr  &   74.14 \\
%Na/Er &  1.24  & Na/Co &   79.77 & Si/Th  &   74.11 \\
%Ca/K  &  1.22  & Rb/La &   78.26 & P/Ni   &   74.11 \\
%Ti/K  &  1.21  & Na/Cr &   78.23 & Mg/Th  &   73.77 \\
\hline\\
\end{tabular}
\caption{Percentages of variance contribute to the total variance and explained variances by the top 10 logratios (PLRs) in each case, as well as the highest percentages of between-phase variance explained. All columns have percentages in descending order. The sum of the contributed variances in the first column (i) of percentages, for all 231 logratios, is 100\% (the total logratio variance, equal to 0.1132). 
Each explained variance in columns (ii) and (iii) is a separate percentage of the logratio variance, the between-sample variance of 0.1132 in column (ii) and the between-phase variance of 0.0475 in column (iii)}.
\label{Ratios}
\end{table*}
\end{center}

These lists give different orderings of the logratios because each one's focus is different. For contributed variances, the top 10 logratios shown here (i.e., 4.3\% of the 231 PLRs), contain 16.6\% of the total logratio variance.
%33.71\%   (for first 23)
This list is quite different from that of the explained variances.
Na appears in all 10 of the list of contributed variances, whereas it appears in none for the explained variances.
The most frequent element in the explained variances column (ii) is K (in 7 out of 10 logratios), followed by Rb (3 logratios). 
The logratio with the highest explanatory power of the whole data set is Mg/K,  explaining as much as 47.3\% of the total logratio variance.
This means that, given just the values of log(Mg/K), all the logratios in the data set can be explained to the tune of an overall $R^2$ value of 47.3\% in simple linear regressions on the explanatory variable log(Mg/K). 
This shows that log(Mg/K) is strongly correlated,  positively or negatively,  with many logratios,  especially those with higher contributed variances.

In column (iii), for quantifying between-phase variance, K appears in 4 of the logratios and Rb again appears in 3 of them. Rubidium commonly substitutes for K in, for example, phlogopite and feldspars.
Some logratios appear in both lists (ii) and (iii): Mg/K, K/Co, Fe/K and Fe/Rb.  
The explained variances between the phases in column (iii) are higher than those in column (ii) and suggest that it will be quite easy to find a model for predicting the five phases.

It is clear that the three different types of variance in Table 1 are measuring different features in the data, which will be used for answering different research questions. 
 The contributed variances in column (i) and the explained variances in column (ii) hint at which logratios might define subcompositions that contain the essential information in the data (an unsupervised objective), whereas the between-phase explained variances in column (iii) indicate the strength of the logratios for predicting the phases (a supervised objective).
 There is some agreement between columns (ii) and (iii) because, as will be seen in later sections, several logratios that account for differences between the samples, also explain differences between the five phases.
This is not always the case, however, and it can happen that there are strong sample patterns found in an unsupervised learning exercise that do not coincide with a supervised learning objective.

%%%%%%%%%%%%%%%%%%%%%%%%%%%%%%%%%%%%%%
\subsection{The role of amalgamations}
Grouping elements by amalgamation (i.e., summing) is a basic strategy of the geochemical data analyst. 
There are often pre-determined amalgamations that are of substantive interest to the research objective.
For example, one way of looking at the data is through the following three amalgamations involving 19 out of the 22 elements in the data:

\begin{enumerate}
\item mantle contamination amalgamation: Si+Mg+Fe+Cr+Co+Ni+Ti
\item crustal contamination amalgamation: Al+Rb+Na+K+Ga
\item kimberlite magma amalgamation: Nb+La+Th+Zr+P+Er+Yb
\item[]
(unused: Ca, Y, V)
\end{enumerate}

\begin{SCfigure}
\includegraphics[width=9cm]{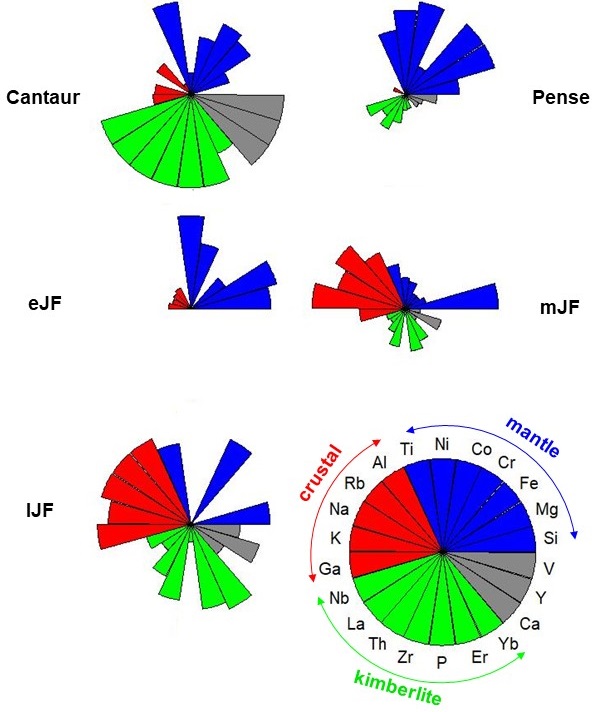}
\caption{Star plots of the five phases according to the three amalgamations of elements. The segment of each element is scaled between the lowest and highest median values of the five phase groups for that element, from the centre to the outer ring respectively. For example, it can be seen that for the earliest Cantuar phase, six of the kimberlite elements, from Nb to Er, have the highest means of these elements, but has the lowest mean for the 
mantle contamination element Si. Three elements are not included in these amalgamations and are coloured gray.}
\label{StarPlot}
\end{SCfigure}

\noindent
These groupings can be shown in a star plot \citep{Becker:88} of the 22 elements and then compared between the five phases -- Figure \ref{StarPlot}. The elements Ca, Y, V were not used as follows. Calcium is present in all three amalgamations and its inclusion does not improve the discrimination between the three processes. Yttrium is primarily associated with the kimberlite fractionation process and could have been included with the kimberlites but is left out and could be considered a possible denominator for the ALR transform. Vanadium appears to occur with both mantle contamination and kimberlite fractionation, and will likely create confusion in the distinction between the phases.

%%%%%%%%%%%%%%%%

Since the three amalgamations define three new parts, they can be examined in a ternary diagram, after closing to obtain a three-part composition.
The mantle amalgamation totally dominates the composition, with means shown in Table \ref{PhaseMeans}.

%%%%%%%%%%%%%%%%
\begin{table}[h]
\begin{tabular}{lccc}
 Phase  & mantle (\%) & crustal (\%) & kimberlite (\%) \\
 \hline
Cantuar & 95.69 & 3.13 &  1.18 \\
Pense & 97.00 & 2.45 &  0.55 \\
eJF & 96.71  & 2.97 & 0.32 \\
mJF & 95.21 & 4.40 & 0.39 \\
lJF & 94.36 & 5.14 & 0.50\\
\hline
\end{tabular}
\caption{Mean percentages for three amalgamations of elements, considered as a new composition with rows summing to 100\%.}
\label{PhaseMeans}
\end{table}

%%%%%%%%%%%%%%%%%%%%%
\begin{figure*}[h!]
\begin{center}
\includegraphics[width=12cm]{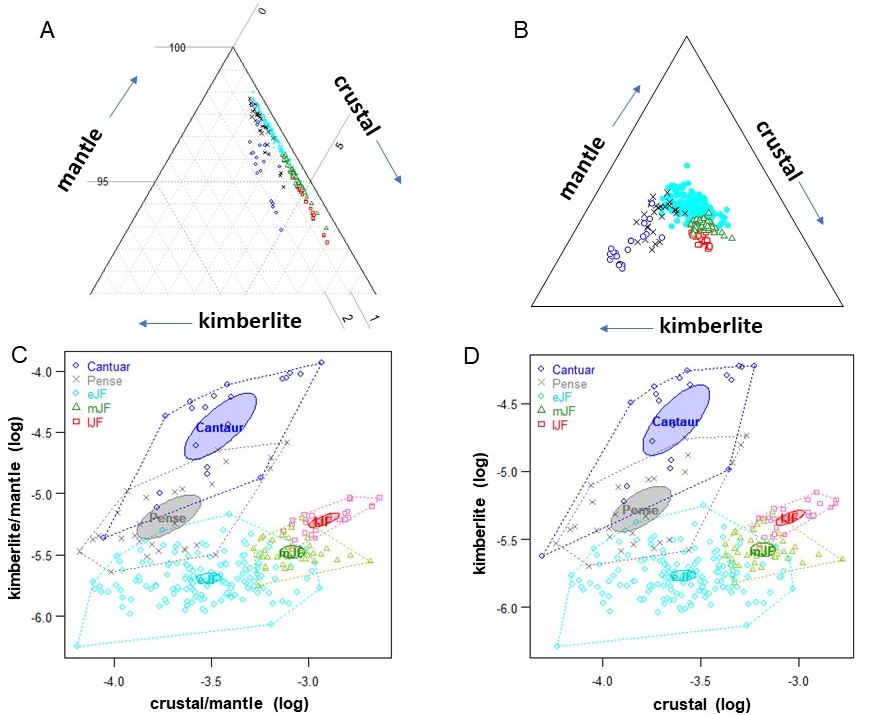}
\caption{A. The top corner of the ternary diagram of the three amalgamations, where all the samples are concentrated. B. The ``centered" ternary diagram of the same data. 
%C. The correspondence analysis of the three-part amalgamation composition, showing the 99\% confidence ellipses for the five phase groups.
C. Scatterplot of two logratios of the three amalgamations, using mantle amalgamation as the denominator, also showing the 99\% confidence ellipses. 
D. Scatterplot of logarithms of the denominators of the logratios in C, showing an almost identical plot. The scales on the axes are shifted by approximately 0.09,  since log(mantle) is close to constant, with mean $-0.092$, where the shift is approximately the negative of this mean.}
\label{Ternary}
\end{center}
\end{figure*}

\medskip
\noindent
Thus, all the points lie in the mantle corner of the ternary diagram, and just that corner is shown in Figure \ref{Ternary}A.
The so-called ``centered" ternary diagram in Figure \ref{Ternary}B centers the mean of all the points at the middle of the triangle, by dividing each amalgamation by its geometric mean and then reclosing the compositions \citep{Eynatten:1, Scholey:1}. This deforms the original scales but the contour lines remain straight, although no longer parallel to the sides of the triangle (see Supplementary Material Section S4, for an example of such contour lines).
The separation of the phases can now be seen more clearly, but the meaning of the distances between the points is not clear.
The differences between phases can be seen in a simple scatterplot of two logratios of the amalgamations, with a much simpler interpretation, -- see Figure \ref{Ternary}C. 
These two logratios form an additive logratio transformation of the amalgamations since they have the same denominator. The choice of denominator can be decided according to the variances of the logarithm of the elements compositional values, which in this case  are:

\smallskip
var(log(mantle amalgamation)) =  0.000145 

var(log(crustal amalgamation)) = 0.107664

var(log(kimberlite amalgamation)) = 0.157917

\smallskip
\noindent
Since log(A/C) = log(A) -- log(C) and log(B/C) = log(B) -- log(C), a very small variance of log(C), i.e. log(C) being almost constant, implies that log(A/C) and log(B/C)) are, up to the subtraction of an almost constant amount, close to log(A) and log(B).  
Hence, it is advantageous to choose the mantle amalgamation as the denominator, due to its log-transform having very low variance. 
Figure \ref{Ternary}D then shows the scatterplot of log(kimberlite amalgamation) and log(crustal amalgamation), which is almost identical to Figure \ref{Ternary}C, apart from the slight shifting of the scales on the two axes.

%The fact that mantle amalgamation is the least discriminating and can also be deduced from the CA biplot in Figure \ref{Ternary}C. 
%This biplot uses the contribution scaling in CA \citep{Greenacre:13}, where the amalgamations most distant from the origin make the highest contributions to the solution.
%Thus, it is really kimberlite and crustal amalgamations that are pulling the samples apart in this biplot, rather than mantle amalgamation.
From Figures \ref{Ternary}C and \ref{Ternary}D, it is clear that higher percentages of the kimberlite fractionation are associated with the earlier Cantuar and Pense eruptions, and increasing percentages of crustal contamination follow the eJF--mJF--lJF temporal sequence of the Joli Fou phases.   
 
In Section S3 of the Supplementary Material, Figure S2 shows the star plots of two other attempts at defining the three amalgamations. 
In terms of discriminating between the five kimberlite phases according to the three amalgamations, the definition in Figure \ref{StarPlot} is the most successful, judged by the percentage of explained variance between phases.

Technically, the logratios within each amalgamation do improve the explanation of the phase differences, with the biggest improvement made by the logratios in the mantle amalgamation (this result is obtained using multinomial logistic regression, described in Section 3.5) . It is not that the structures within the amalgamations are not relevant, it is rather that the amalgamations are structured to represent similar processes and already explain the phase differences well (Figure \ref{Ternary}C), using just three amalgamated parts that are easily interpretable. The group of mantle-associated elements represent a unique mineralogy, which makes amalgamation particularly effective. Adding additional logratios complicates the interpretation but can certainly be explored for their substantive usefulness.

%%%%%%%%%%%%%%%%%%%%%%%%%%%%%%%%%%%%%%%%%%%%%%%%
\subsection{Multivariate analysis: unsupervised}
Unsupervised multivariate analysis attempts to identify structure between all the logratios in the form of underlying dimensions, often called factors, or groups of samples or groups of variables, also called clusters. 
Cluster analysis is the easiest to understand and interpret, since forming groups of objects is a natural objective in the sciences.
There are generally two types of clustering algorithms: hierarchical clustering, usually applied to less than approximately 100--150 objects in order to show the results in a dendrogram (although we will show it for the 270 samples), and non-hierarchical clustering, usually applied to large to very large sets of objects where the objective is just to partition the objects into internally homogenous groups, called clusters. 
The data set here of 270 samples is already grouped into the five phases, but how would a clustering algorithm group the samples simply based on their geochemical compositions, without any external knowledge?

In order to perform clustering, a distance function between objects is required, and there are several possibilities:
\begin{enumerate} 
\item Transform the compositional data to CLRs and then use Euclidean distance. This is clustering on the (unweighted) logratio distances. 
%\item Transform the compositional data to weighted CLRs and then use Euclidean distance. This is clustering on the weighted logratio distances.
\item As above, transform the compositional data to CLRs but then alternatively use nonhierarchical k-means clustering, which also uses Euclidean distance. 
\item Transform the compositional data to ALRs, either statistically optimal ones, or based on geochemical justification, and then use Euclidean distance; here statistical optimality is defined as producing the geometry closest to the geometry of all logratios  \citep{GreenacreMartinezBlasco:21}. 
\end{enumerate}

\noindent
There is a fourth option, namely to use the chi-square distance inherent in correspondence analysis, which is  a useful alternative for compositional data with zeros -- see the discussion in Section 4.
Here we show the results of the hierarchical clustering of the first option, with some results for the third one and the chi-square option given in Supplementary Material Section S5.

Figure \ref{HierClust} shows the result of the first clustering, using function \texttt{hclust} in \textsf{R}, with horizontal plotting of the dendrogram using the \textsf{R} package \texttt{ape}. 
The endpoints show symbols according to the five phases.
A five-cluster solution looks feasible since there is a large gap between the nodes on either side of the cutpoint shown as a dashed line.
Although the objective here has not been to classify the phases, the association between the five clusters and the five phases can be summarized in Table \ref{CrossTable1}.
Assigning clusters to the most frequent phases gives 66.7\% correct assignments, but there is no assignment to Pense or mJF.

\medskip

%\begin{SCfigure}
\begin{figure}
 \begin{center}
\includegraphics[width=10.7cm]{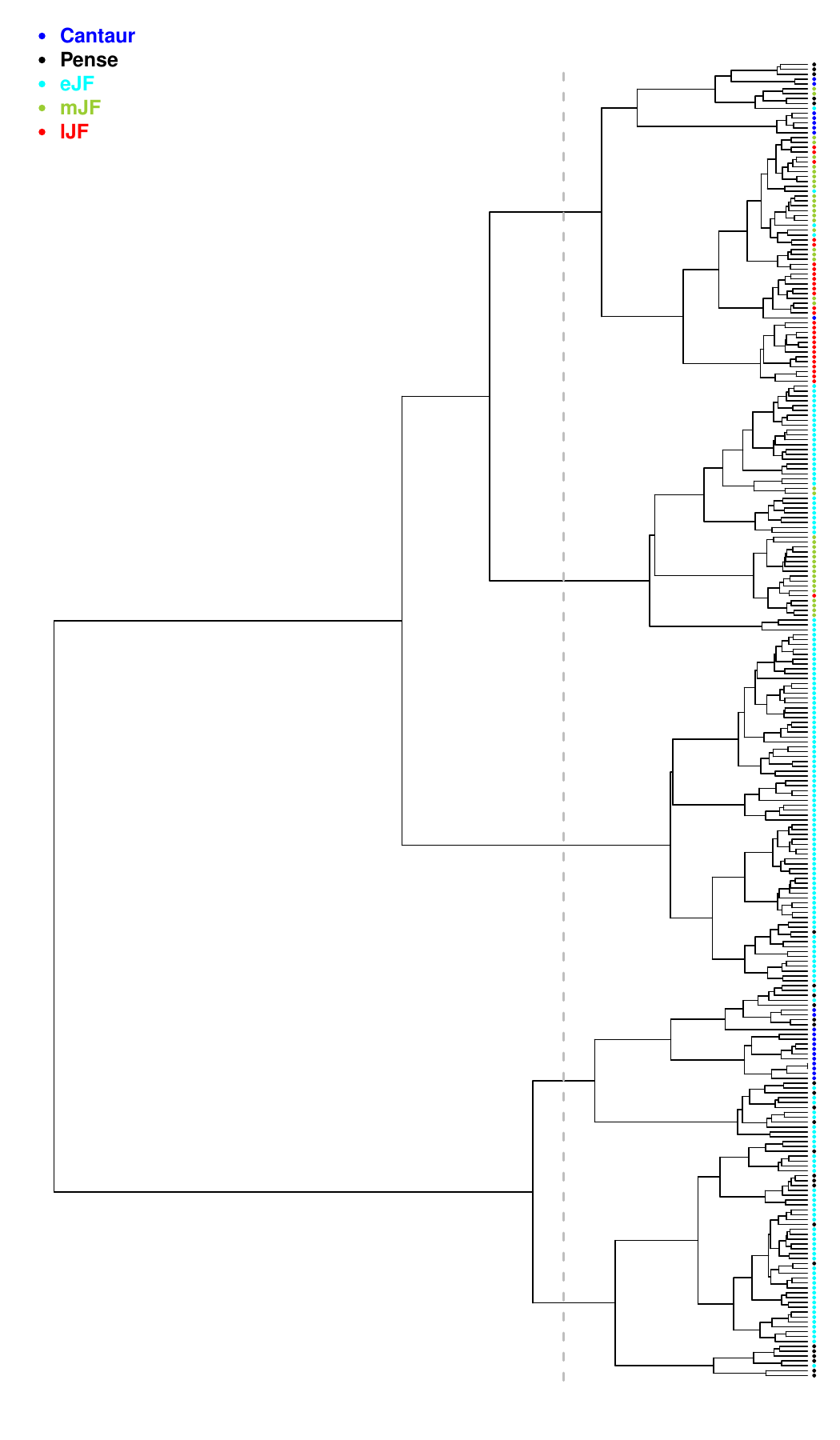}
\caption{Ward hierarchical clustering of logratio distances between samples, indicating the cutpoint for a five-cluster solution.}
\label{HierClust}
\end{center}
\end{figure}
%\end{SCfigure}

\begin{table}[h]
\begin{tabular}{crrrrr}
Cluster &  Cantuar & Pense & \quad eJF & \quad mJF & \quad lJF \\
\hline
  1   &     0  &    0 &  32 &  18 &   1 \\
  2   &    13  &    9 &  10 &   0 &   0 \\
  3   &     0  &   12 &  37 &   0 &   0 \\
  4   &     0  &    1 &  71 &   0 &   0 \\
  5   &     8 &     5 &   4 &  22 &  27 \\
\hline
\end{tabular}
\caption{Cross-tabulation of five clusters of the samples, obtained from hierarchical Ward clustering (Figure \ref{HierClust}), versus their phase classes.}
\label{CrossTable1}
\end{table}
\medskip
\noindent

\begin{figure*}[htbp]
\begin{center}
\includegraphics[width=9.5cm]{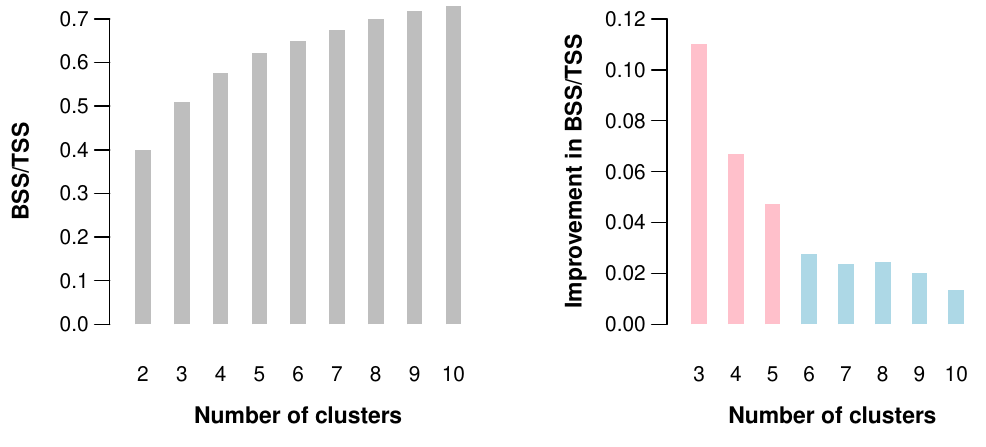}
\caption{On the left, the proportion of between-cluster sum of squares (BSS) out of total sum of squares (TSS) for increasing numbers of clusters in k-means non-hierarchical clustering. On the right, the improvements are shown: for example, for three clusters, the improvement gained in going from the 2-cluster solution to the 3-cluster one is equal to 0.109 (10.9\%). This right hand plot is read like a scree plot in PCA and it is clear that the 5-cluster solution is preferred, since the gain from the 4-cluster solution to the 5-cluster one is large, and after that the gains are less and quite similar.}
\label{NonHierClust}
\end{center}
\end{figure*}

As an alternative to hierarchical clustering, non-hierarchical clustering using the k-means algorithm is provided by the function \texttt{kmeans()} in \textsf{R}.
This function requires the number of clusters to be pre-specified, so the algorithm is applied using increasing numbers of clusters, and each time the ratio of between-cluster sum of squares (BSS) is computed relative to total sum of squares (TSS).
In Figure \ref{NonHierClust} this ratio increases with increasing number of clusters, as expected, but it is easier to judge the improvement by plotting the increases with each additional cluster, shown in the right hand plot. 
This plot is read exactly like a scree plot in a PCA, where the ``broken-stick" (or ``elbow") visual diagnostic suggests that the 5-cluster solution is the one of choice, which coincides with the number of phases.

The results of the 5-cluster solution are given in Table \ref{CrossTable2}.
There are 67.8\% correct assignments, representing a slight improvement over the first hierarchical clustering reported above (Figure \ref{HierClust}), which uses the same logratio distances. 
The results can be slightly improved by investigating all the k-means solutions for reduced-dimensional versions of the logratio-transformed data -- see \cite{DingHe:1} and references therein.
It turns out that the two-dimensional solution, using the first two principal components, is the most successful,
%with a Cramer's V measure of 0.601 and 
with a percentage of correct assignments equal to 71.1\%.

\begin{table}[ht]
\begin{tabular}{crrrrr}
Cluster &  Cantuar & Pense & \quad eJF & \quad mJF & \quad lJF \\
\hline
  1   &     0  &    0 &  35 &  32 &   4 \\
  2   &    13  &   11 &   8 &   0 &   0 \\
  3   &     0  &    0 &  70 &   0 &   0 \\
  4   &     8  &    4 &   0 &   0 &  24 \\
  5   &     0 &     2 &  41 &   0 &   0 \\
\hline
\end{tabular}
\caption{Cross-tabulation of five clusters of the samples, obtained from nonhierarchical k-means clustering, versus their phase classes.}
\label{CrossTable2}
\end{table}
%Cramer's V is equal to 0.555, with 

It should be stressed again that the objective here is not to ``predict" the phases, but to investigate the association between the natural clustering in the data and the given classification into five phases -- the predictive objective will be dealt with in the next section.

\begin{figure*}[htbp]
\begin{center}
\includegraphics[width=11cm]{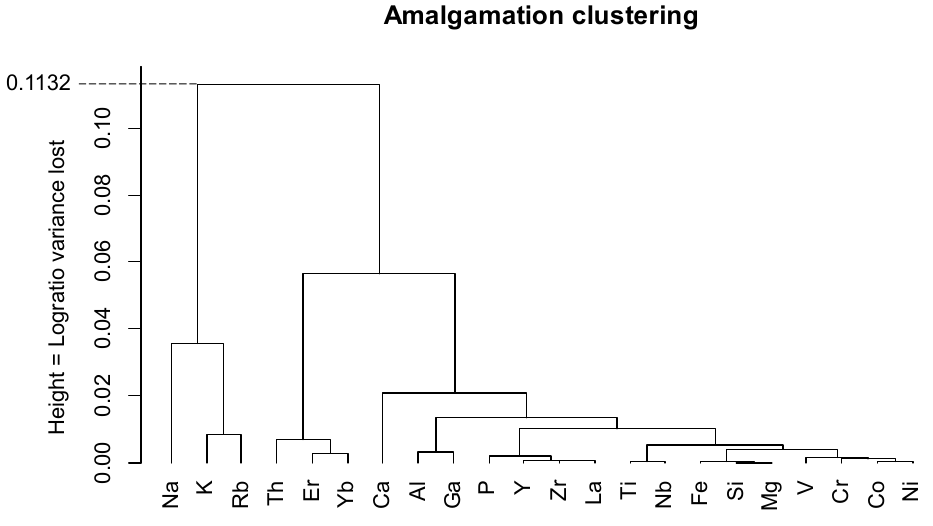}
\caption{Amalgamation clustering of the kimberlite data. The algorithm proceeds from no variance lost (zero, because all the individual elements explain all of their own logratio variance) to the total variance of 0.1132 lost (when all the elements are merged, and they explain none of their own variance).}
\label{AmalgClust}
\end{center}
\end{figure*}

To close the treatment of cluster analysis, the clustering of the elements themselves is considered.
The most natural way to merge elements is to simply amalgamate (i.e., sum) their compositional values.
The algorithm is given by \cite{Greenacre:20} and is based on explained logratio variance.
When two elements are amalgamated, then the reduced data set, with one less element, explains less than the total logratio variance, that is, some variance is unexplained, or lost, in the process.
The ``closest" two elements are those which, if amalgamated, cause the least reduction in explained logratio variance. 
Each step of the algorithm proceeds in this way, always amalgamating the elements (or ones already amalgamated), which lead to the least reduction in explained variance. 
At the end of the algorithm, when all elements are amalgamated, all the variance is lost, so the highest value on the dendrogram is the total logratio variance, 0.1132.
Figure \ref{AmalgClust} illustrates two separate concepts of interest. (1) elements with similar charge and ionic radius: e.g., Na, K-Rb; Er, Yb; Al-Ga; and (2) mineralogical control: e.g., Fe-Si-Mg (olivine) and V-Cr-Co-Ni substituting in olivine as minor/trace elements, and Ti-Nb in perovskite and or ilmenite.

The approach that is complementary to clustering for unsupervised analysis is to do a PCA, and again there are alternative ways to do this, in parallel with the options that were investigated in the clustering above.
The first way is to do the PCA on the CLRs, which is equivalent to the PCA of all pairwise logratios \citep{AitchisonGreenacre:02}, thus called logratio analysis (LRA) \citep{Greenacre:18, Greenacre:21}.
The second way, if a set of ALRs is preferred from the start, is to perform PCA on the chosen set of ALRs. This option is demonstrated using reference element Zr, identified as giving the set of ALRs closest to the geometry of all logratios -- this denominator is found based on the Procrustes correlation \citep{Greenacre:18}, using the function \texttt{FINDALR} in the \texttt{easyCODA} package.
The complete set of 22 Procrustes correlations, using each of the elements in turn as the reference, is given in Table S1 of the Supplementary Material (Section S6).
The higher the Procrustes correlation, the closer the geometry of the ALR transforms is to the exact logratio geometry -- using Zr as the denominator gives a maximum Procrustes correlation of 0.977. 
It is worth noting that Zr, along with La and Y account for a minimal amount of variance and are not associated with elements of similar charge or mineralogical control within the mafic rocks (olivine) -- this makes such elements suitable for an ALR transformation. 
Furthermore, Zr is a highly immobile element and is resistant to alteration and weathering processes. 
 Once again, an alternative option is to use correspondence analysis (CA) on the original compositions, preferably with a prior power transformation \citep{Greenacre2023chiPower} -- the results are given in Supplementary Material Section S7.
These alternative approaches all result in a compositional biplot where the samples as well as their phase group means in the two-dimensional reduction can be shown. 
%In addition, convex hulls around the phase group samples can be shown, as well as confidence ellipses for indicating precision of the phase means. 

\begin{figure*}[htbp]
\begin{center}
\includegraphics[width=11cm]{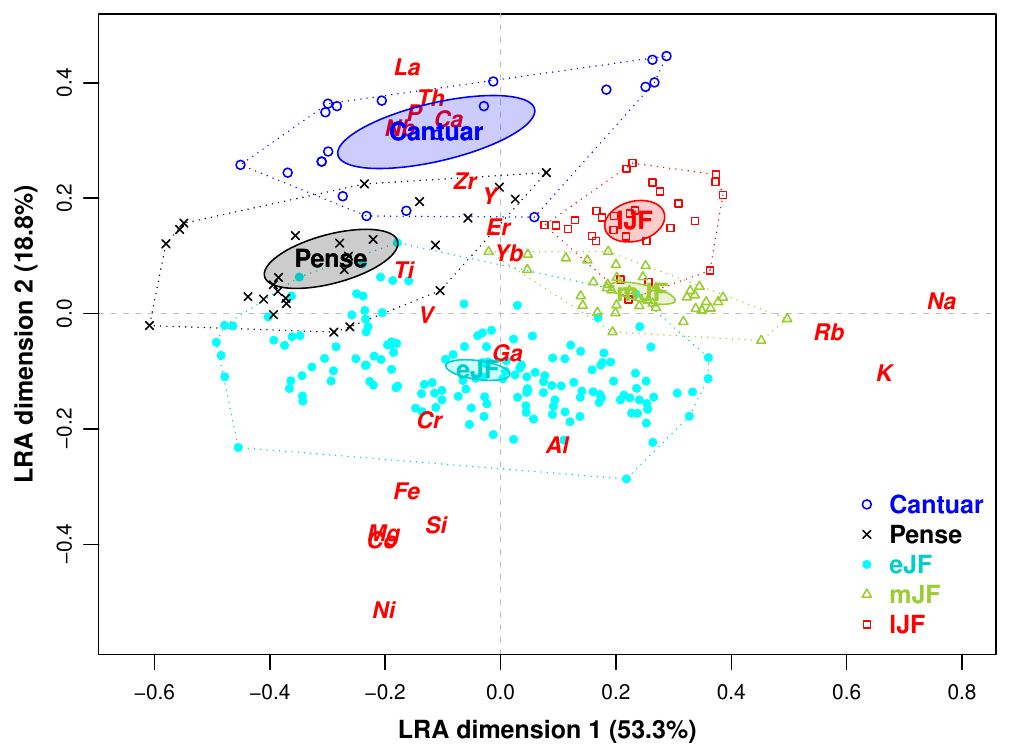}
\caption{Logratio analysis of the kimberlite data. Variance explained in this two-dimensional solution is 72.1\%.}
\label{BiplotLRA}
\end{center}
\end{figure*}

Figure \ref{BiplotLRA} shows the biplot of the logratio analysis of the kimberlite data -- in fact, it is in this planar solution that the sample points have been clustered by k-means in the two-dimensional reduced data set mentioned previously.
Note that the interpretation of the elements, displayed as projected CLRs, should not be on the positions of the elements themselves, although the usual way interpreting PCA would suggest this way.
It is rather the connections between pairs of elements, representing the pairwise logratios, that should be interpreted.
For example, to interpret why the Cantuar samples and the eJF ones are separated vertically, one has to look for pairs of elements that connect vertically: for example, La and Si, where the direction of an arrow from Si to La would represent the logratio log(La/Si).
The fact that there are some elements such as La, Th, P, Ca and Nb close together at the top of Figure \ref{BiplotLRA}, means that they will make similar connections with any other element, for example Ni, representing logratios log(La/Ni), log(Th/Ni), log(P/Ni), etc., which will be logratios all separating Cantuar from eJF. 
The phases are well separated in the PCA plane, even though no information has been given to this statistical method about the phase classes.

The dispersions of the samples in their phase groups are indicated in two ways: first, by convex hulls (dotted lines) that enclose all the sample points of the respective phases, and secondly, by 99.5\% confidence ellipses around the phase mean points, where the positions of the means are labelled by the phase names.
The level of confidence of 99.5\% is based on the conservative Bonferroni correction of the significance level of 0.05.
Since there are 10 pairwise comparisons between the five phases, the significance level is 0.05/10 = 0.005, hence the use of a confidence level of $1-0.005 = 0.995$. 
The large separation of all five confidence ellipses implies that the five phases are all significantly different pairwise, at an overall significance level less than 0.05 \citep{Greenacre:16b}.

\begin{figure*}[htbp]
\begin{center}
\includegraphics[width=11cm]{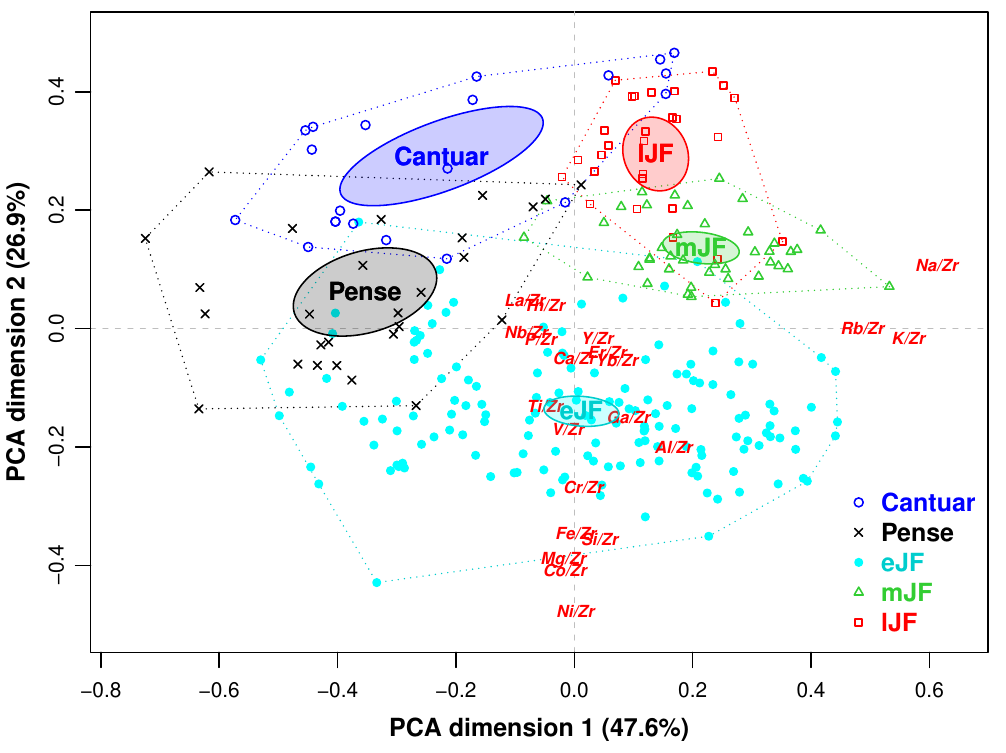}
\caption{Principal component analysis of the ALR-transformed kimberlite data, using Zr as the reference (i.e., denominator) element. Variance explained in this two-dimensional solution is 74.5\%.}
\label{BiplotPCA}
\end{center}
\end{figure*}

Figure \ref{BiplotPCA}, shows the PCA of the 21 ALRs using Zr as the denominator element. So, seeing Zr's position in Figure \ref{BiplotLRA}, we will expect logratios such as log(Ni/Zr), log(Mg/Zr), log(Si/Zr), etc., to be correlated, and separate out eJF.
This is indeed the case in Figure \ref{BiplotPCA}, where those ratios are pointing downwards.  
In this PCA the logratios are regular variables, so the interpretation here is standard.
Again the phases are well separated, very similar to Figure \ref{BiplotLRA}.
This is not surprising, since the denominator element Zr has been chosen because it gives a set of ALRs very close to the logratio geometry.

%%%%%%%%%%%%%%%%%%%%%

 In the previous analyses, all the elements have been used, but it could be that certain elements or their logratios play a very minor role in process discovery, and can be excluded.
This can be investigated, for example, by searching for subsets of logratios that explain the logratio geometry the most -- see \cite{Greenacre:19}.
Furthermore, since there are always several logratios that compete closely in this search, expert knowledge can be incorporated to select certain logratios that have more geochemical meaning.
This strategy has been successful in two studies, in biochemistry \citep{Graeve:20} and archaeology \citep{Wood:21}, leading to a reduction in the number of compositional parts that need to be considered and a consequent simplification of the results.

%%%%%%%%%%%%%%%%%%%%%%%%%%%%%%%%%%%%%%%%%%%%%%%%%%%%%%%%%%%%%%%%%%%%%
\subsection{Multivariate analysis: supervised}
While unsupervised multivariate analysis attempts to identify structure in the individual samples, with no external objective, supervised analysis is focused on using the multivariate compositions to explain or predict a response variable.
Alternatively, depending on the research objective, external explanatory variables can be used to explain the compositions themselves.
In the present data set, the only additional variable that is available is the categorical kimberlite phase variable, which can be regarded as a response to the geochemical compositions.
It is possible to model phase group as a function of the compositions in different ways, using methods such as analysis of variance, logistic regression or classification trees.

\begin{figure*}[h]
\begin{center}
\includegraphics[width=13cm]{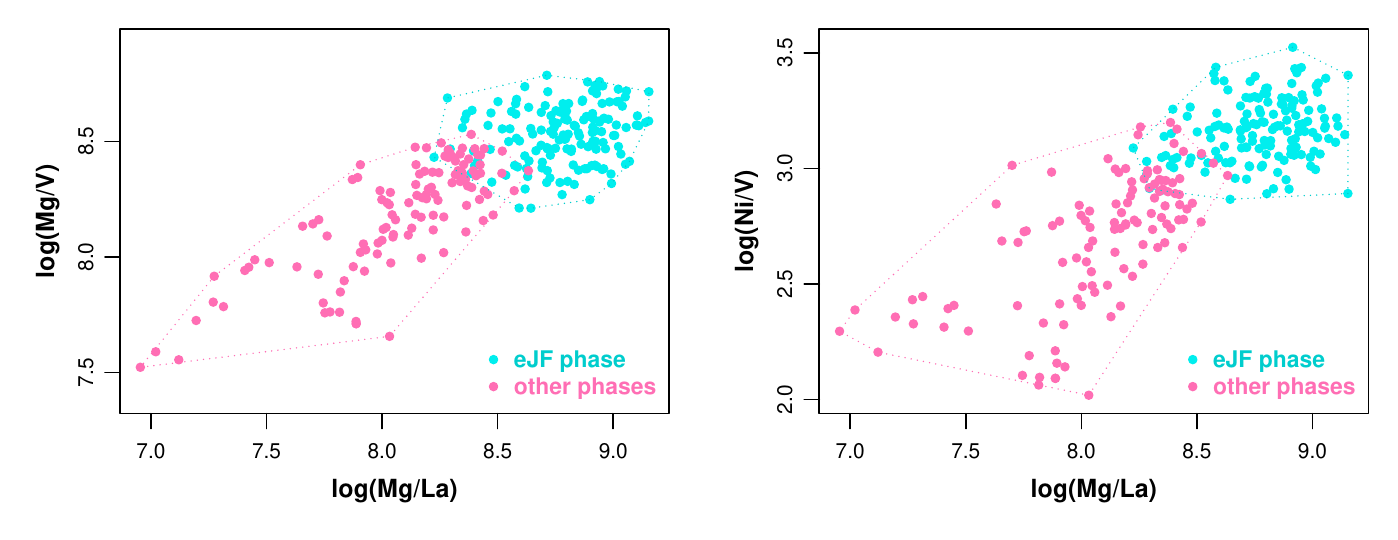}
\caption{Scatterplots of two logratio predictors of eJF and the other phases, first with the best pair of logratios (left) and second with the best pair of non-overlapping logratios (right).}
\label{LRpreds}
\end{center}
\end{figure*}

As a simple start, a two-group response is considered to predict early Joli-Fou (eJF), versus all the other phases, since eJF is the phase originally singled out as important for the discovery of macro-diamonds  \citep {Harvey:1, Grunsky:30}.  
\cite{Coenders:22} detail three alternative methods for choosing pairwise logratios as predictors of a categorical or continuous response. 
It is a stepwise search and here the most conservative stopping rule is chosen, which is the default in the \textsf{R} function \texttt{STEPR} in package \texttt{easyCODA}, based on the Bonferroni criterion.
The three alternative methods are (1) select any of the logratios (of which there are 213); (2) select from the restricted set of non-overlapping logratios, which have no elements in common; (3) select additive logratios (ALRs) with the same denominator.
In the present application, all alternatives choose only two logratios, and methods (1) and (3) coincide, so only two solutions need to be presented.
Using method 1, two logratios log(Mg/La) and log(Mg/V) are chosen, with element Mg overlapping. 
Using method 2, log(Mg/La) is still chosen first but the non-overlapping  log(Ni/V) is chosen as the second best logratio predictor.
Scatterplots of the pairs of chosen logratios are shown in Figure \ref{LRpreds}.

The logistic regressions that are used in the variable selection process to model the two  kimberlite phase groups (eJF phase versus other phases) as functions of the logratio predictors given above, respectively give 252 and 253 (out of 270, 93.3\% and 93.7\%, respectively) correct predictions of the two phase groups, which is an excellent result, as was expected from earlier considerations in Section 3.3.
However, this is an over-optimistic result of the predictive success, so it is better to perform a cross-validation on these models to ascertain better their prediction accuracy.
A 10-fold cross-validation \citep{Hastie:09} was performed and the results were only slightly less successful: 250 and 252 correct predictions respectively for the two types of variable selection.
About 93\% prediction accuracy can thus be expected, using just two logratios, although it should be remembered that the cross-validation is done internally on the whole data set, not on a separate training set and then applied to a ``hold-out" test set that is separate from the training set (see the results of the random forests algorithm reported below).

The results of the modelling can be shown algebraically or graphically.
The two estimated models for the log-odds-ratio, logit($p$), where $p$ is the probability of phase eJF, are as follows:
\begin{align}
  \textrm{logit}(p) &= -228.81 + 14.02 \log(\textrm{Mg/La}) + 13.15 \log(\textrm{Mg/V}) \label{LogratioModel1} \\ 
  \textrm{logit}(p) &= -119.10 + 10.20 \log(\textrm{Mg/La}) + 11.03 \log(\textrm{Ni/V}) \label{LogratioModel2}
\end{align}
The interpretation of the regression coefficients in these two models is different, where only the second equation, with its non-overlapping logratios, is able to be interpreted in the usual way.
However, the linear combinations of logratios in both equations can be reduced to log-contrasts of the elements involved \citep{Coenders:20, Coenders:22}.
For example, for the first model (\ref{LogratioModel1}), the log-contrast is defined for three elements with coefficients summing to zero, as follows: 
\begin{equation}
    27.17 \log(\textrm{Mg}) -14.02  \log(\textrm{La}) - 13.15 \log(\textrm{V}) \nonumber
\end{equation}

\begin{SCfigure}
\includegraphics[width=8cm]{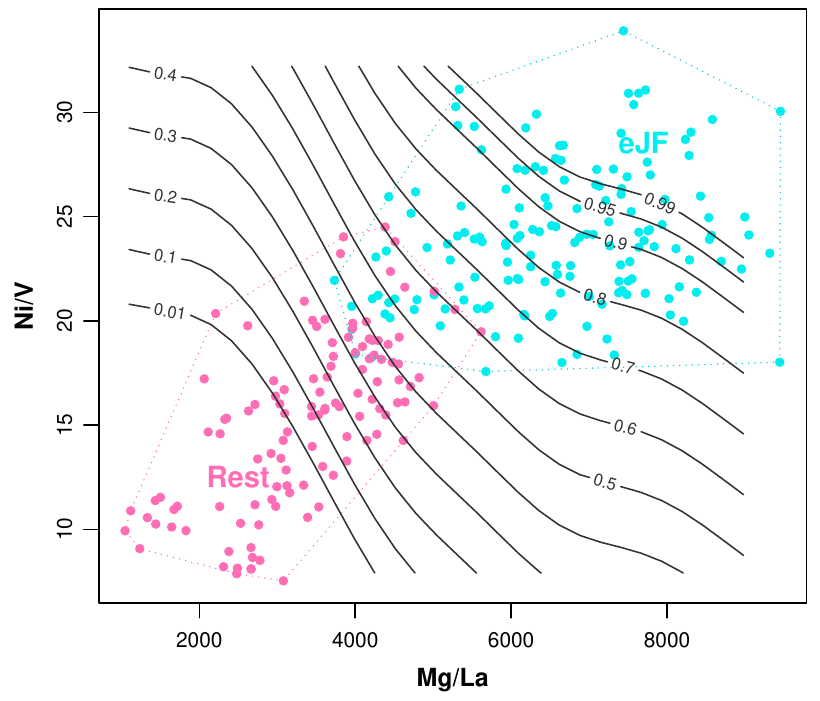}
\caption{Right hand plot in Figure \ref{LRpreds}, now plotted to the scale of the ratios (i.e., not log-transformed), along with probability contours for predicting eJF, back-transformed from the logistic regression model (\ref{LogratioModel1}). The misclassified samples can be seen on either side of the 0.5 predicted probability contour level.}
\label{Contour}
\end{SCfigure}

The equations can also be visualized in a contour plot of the predicted probabilities $p$ on two dimensions, using the the ratios themselves as coordinate axes.
That is, these are contours of the back-transformed probabilities (from the logit values) of eJF as a function of the ratio predictors with no log-transformation -- see Figure \ref{Contour}, based on equation (\ref{LogratioModel1}).
Classification is made on either side of the 0.5 probability contour. Above the 0.5 contour, eJF is classified, and other phases below. The 18 misclassified samples can then be seen just below and above the 0.5 contour.

\begin{SCfigure}
%\begin{center}
\includegraphics[width=8cm]{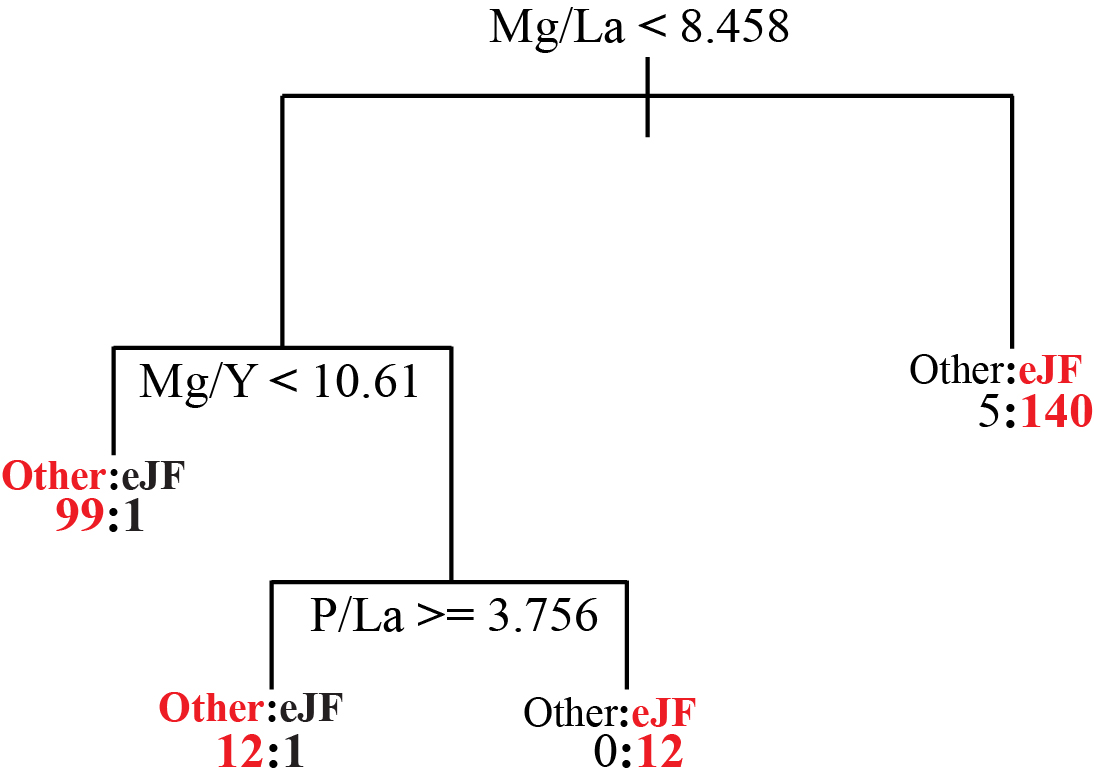}
\caption{Classification tree predicting eJF or other phases. At each split the condition indicates which samples go to the left. This tree gives 263 correct predictions (out of 270, i.e. 97.4\%), but a cross-validation estimates 251, which is a precision of 93.0\%.}
\label{Tree}
%\end{center}
\end{SCfigure}

Usually, a classification tree approach improves over the result of a logistic regression, and the present case shown in Figure \ref{Tree} is no exception --- 263 out of 270 samples (97.4\%) are correctly predicted. 
But that improvement is reduced for the cross-validation, where a prediction accuracy of 251 was achieved (93.0\%), similar to the logistic regressions. 
Figure \ref{Tree} uses all the data, and shows the splits in terms of the ratio values.
To construct the tree classifier, no logarithmic transformation is needed, since any monotonic version of the ratios does not change the result. 
It can be seen that Mg/La is still the most important ratio, but other ratios are chosen here that do not involve element V.  

Based on Figure \ref{Tree}, eJF and Other are predicted according to the following rules:
\begin{itemize}
    \item Predict eJF if (a) Mg/La is greater than (or equal to) 8.458; 145 samples satisfy this condition, of which 140 are eJF, and 5 are misclassified as Other.
    \item Predict eJF if (a) Mg/La is less than 8.458, and Mg/Y is greater than or equal to 10.61 and P/La is less than 3.756; 12 samples satisfy this set of conditions and they are all eJF.
    \item Otherwise predict Other (i.e., not eJF); of the remaining 113 samples, 111 are Other, and thus correctly predicted.
\end{itemize}

If all five phases are required to be predicted, either a multinomial logistic model has to be fitted, or a classification tree has to be estimated.
For the multinomial logistic model, one phase has to be defined as a reference for the other four, and the earliest phase, Cantuar, is chosen. This choice does not affect the phase predictions, only the coefficients and their standard errors for each of the four models.
The same Bonferroni criterion of \cite{Coenders:22} is used, with the number of parameters estimated being $m = 4p$, where $p$ is the number of logratio predictors.
Using this criterion, three logratios are chosen: log(Si/Zr), log(P/Rb) and log(Si/Ni).
The estimated models are given in Table S4 of Section S8 of the Supplementary Material.
The  classification accuracy of this multinomial logistic model is shown in Table \ref{CrossTable3}, with 19 samples misclassified, that is 251 predicted correctly (93.0\%).

\begin{table}[h]
\begin{tabular}{lrrrrr}
Predicted & \multicolumn{5}{c}{True phase}\\[-0.2em]
 phase &  Cantuar & Pense & \quad eJF & \quad mJF & \quad lJF \\
\hline
  Cantuar   &    20  &    1 &   0 &   0 &   0 \\
  Pense     &     0  &   22 &   1 &   0 &   0 \\
  eJF       &     0  &    4 & 151 &   2 &   1 \\
  mJF       &     0  &    0 &   2 &  35 &   4 \\
  lJF       &     1  &    0 &   0 &   3 &   23 \\
\hline
\end{tabular}
\caption{Cross-tabulation of predicted phases, obtained from multinomial logistic modelling, versus the actual phase classes. Percentage of correct predictions is 93.0\%.}
\label{CrossTable3}
\end{table}

\begin{figure}[h]
\begin{center}
\includegraphics[width=12cm]{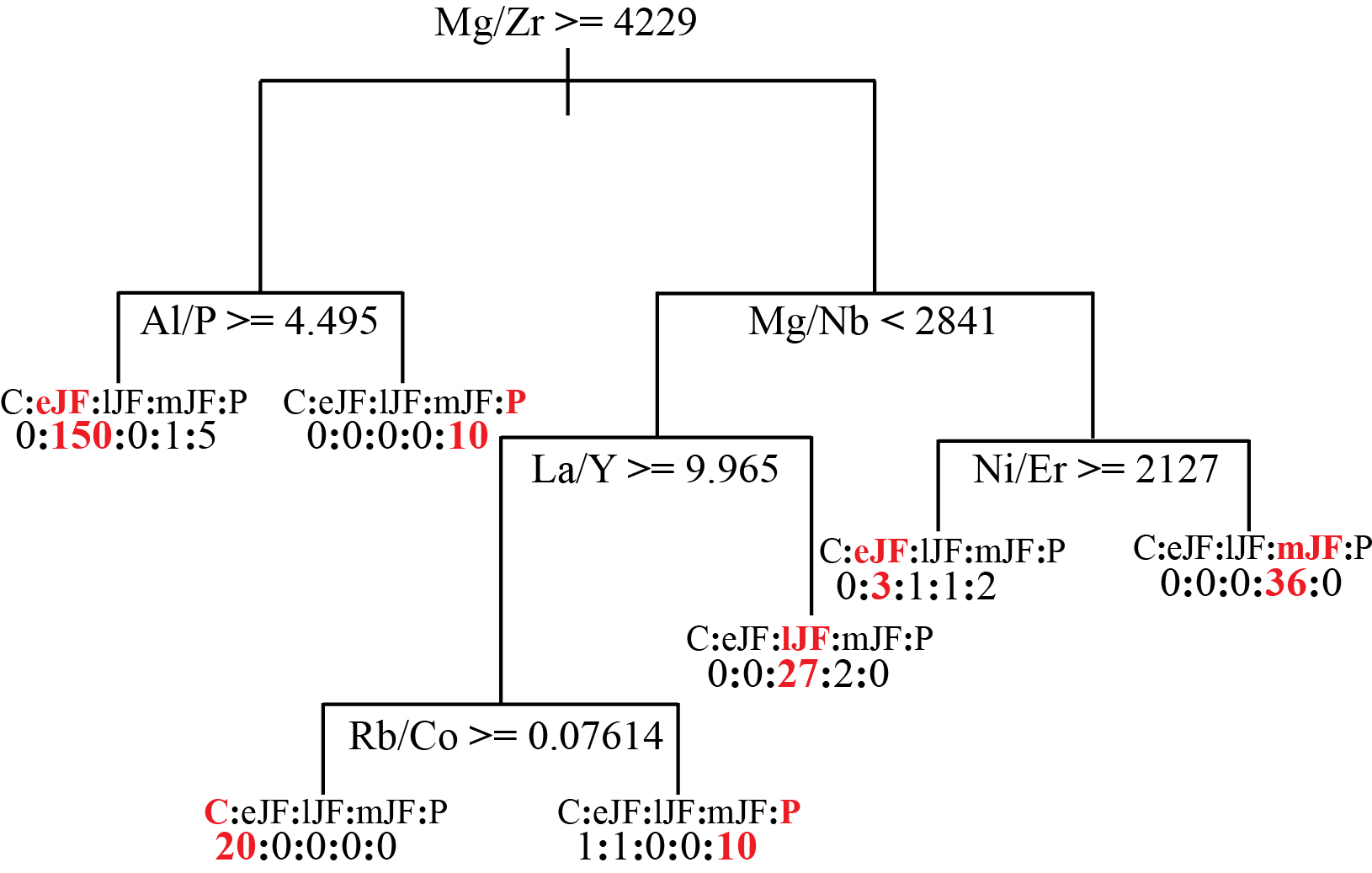}
\caption{Classification tree predicting all five phases. At each split the condition indicates which samples go to the left. This tree gives 256 correct predictions (out of 270, i.e. 94.8\%), but a cross-validation gives only 233 correct, which is a precision of 86.3\%.}
\label{Tree5}
\end{center}
\end{figure}

\begin{table}[h]
\begin{tabular}{lrrrrr}
Predicted & \multicolumn{5}{c}{True phase}\\[-0.2em]
 phase &  Cantuar & Pense & \quad eJF & \quad mJF & \quad lJF \\
\hline
  Cantuar   &    20  &    0 &   0 &   0 &   0 \\
  Pense     &     1  &   20 &   1 &   0 &   0 \\
  eJF       &     0  &    7 & 153 &   2 &   1 \\
  mJF       &     0  &    0 &   0 &  36 &   0 \\
  lJF       &     0  &    0 &   0 &   2 &  27 \\
\hline
\end{tabular}
\caption{Cross-tabulation of predicted phases, obtained from multinomial logistic modelling, versus the actual phase classes. Percentage of correct predictions is 94.8\%.}
\label{CrossTable4}
\end{table}

Alternatively, the classification tree classifies the five phases as shown in Figure \ref{Tree5}, with classification accuracy given by Table \ref{CrossTable4}, and 14 phase classes misclassified, that is 256 predicted correctly (94.8\%), five more than the multinomial logistic model. 

\begin{SCfigure}[][h!]
\includegraphics[width=8cm]{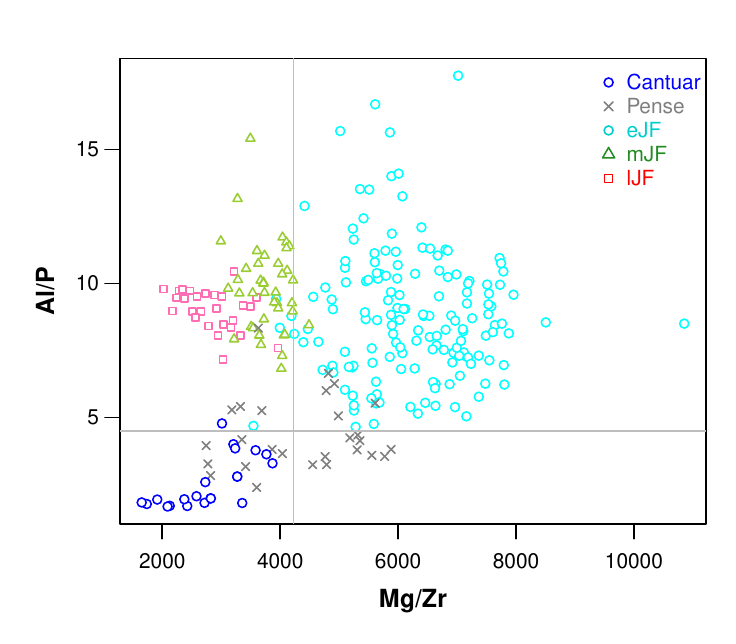}
\caption{Scatterplot of two ratios and the two thresholds, Mg/Zr = 4229 and Al/P = 4.495, above which the classification tree of Figure \ref{Tree5} predicts eJF with high probability (upper right block).
The 10 correctly predicted Pense can also be seen isolated in the bottom right block of the plot, where Mg/Zr $\geq$ 4229 and Al/P $<$ 4.495.}
\label{ScatterPlot2}
\end{SCfigure}

As for Figure 13, each end branch of the tree in Figure \ref{Tree5} can be described by a set of conditions. Here eJF is better classified, mainly according to the left hand branch with the following rule:
\begin{itemize}
  \item Predict eJF if Mg/Zr is greater than or equal to 4229 and Al/P is greater than or equal to 4.495
\end{itemize}
If Mg/Zr and Al/P are plotted in a scatterplot (Figure \ref{ScatterPlot2}), the 150 correctly predicted eJFs, and the 1 misclassified mJF and 5 misclassified Pense samples can be seen.

As before it is necessary to do some type of cross-validation to assess the true predictive accuracy of these two methods.
In this case the multinomial regression does better, with a prediction accuracy of 89.3\% (241 correct out of 270, 29 incorrect predictions.
On the other hand, the classification tree cross-validates to an accuracy of 86.3\% (233 correct out of 270, 37 incorrect predictions).
Notice again that this is a cross-validation on the complete data set, not the prediction performed on a separate hold-out sample, as is done in the random forest prediction below..

It is instructive to follow up some samples that are misclassified and to see the reasons why.
For example, there is one sample (number 115) in the data set that is predicted to be Pense, but is actually eJF, predicted as Pense by both the multinomial logistic model and the classification tree.
The count of 1 can be seen in the two corresponding misclassification tables, Tables 5 and 6 (row Pense, column eJF), and is the 1 in the lowest terminal node of Figure \ref{Tree5} with 12 samples, where Pense is predicted due to the 10 Pense samples, with two misclassifications, one being Cantuar and the other being eJF, the latter being the sample in question here. 
Looking at its relevant logratios, Mg/Zr = 3547, due to its lower Mg and higher Zr compared to the values in general for the eJF phase. This has taken this sample to the right hand side of the top split in Figure \ref{Tree5}, and is the reason why it is misclassified. If it had gone to the left, as most of the eJF samples have done, its value of 4.69 for Al/P would have taken it to the end node at extreme left where almost all of the eJF samples were correctly classified.

Finally, a random forest classification produces no single model but serves only as a prediction algorithm. 
The data set is divided randomly into a training set, consisting of 181 samples and a test hold-out set consisting of 89 samples -- the selection was done using stratified random sampling with the objective of having 2/3 of the samples in the training set and 1/3 in the test set, with similar distributions of the phase groups in each set.
The random forest algorithm, using function \texttt{randomForest()} in the \textsf{R} package of the same name \citep{Liaw:02} produces an improved result, predicting 76 of the 89 samples correctly (i.e., 88.8\% correct), where it should be remembered that the test set is separate from the training set on which the prediction model has been developed. 
The out-of-bag (OOB) classification prediction \citep{Hastie:09} on omitted samples during the training of the model with bootstrap samples is 92.8\%, doing much better than the cross-validation of the single tree classifier in Figure \ref{Tree5}.

%%%%%%%%%%%%%%%%%%%%%%%%%%%%%%%%%%%%%%%%%%%%%%%%%%%%%%%%%%%%%%%%%%%%%
\section{Discussion and concluding remarks}
The field of compositional data analysis has expanded to unanticipated proportions since Aitchison's involvement in the subject in the 1980s. The logratio concept has opened up an entirely new way of evaluating compositional data without the risk of spurious results.

Our proposed GeoCoDA workflow does not include so-called isometric logratio (ILR) ``balances"  (renamed ``orthonormal logratios" by some authors), which are in our experience too complicated and difficult to interpret \citep{Aitchison:08, GreenacreGrunskyBaconShone:20, GreenacreEtAl:23}.
 An ILR balance is proportional to the logratio of geometric means of $J_1$ and $J_2$ parts respectively in the numerator and denominator, which is identical to the average of $J_1 \times J_2$ PLRs.
This is interpretable when there are very few parts involved, for example $J_1=1$ and $J_2=2$, which is the average of two PLRs, otherwise it is not clear that more complex balances necessarily include PLRs that are related to the research objective. 
In a comprehensive reappraisal of Aitchison's legacy after 40 years since his paper to the Royal Statistical Society \citep{Aitchison:82}, \cite{GreenacreEtAl:23} have concluded that ILR balances are not a necessary prerequisite for good practice in CoDA, in spite of many claims to the contrary that they are the required transformations to use.
Alternative simpler transformations such as the CLR or ALR yield substantively equivalent results, with an easier and clearer interpretation.
 CLRs, ALRs and PLRs are all convenient for variable selection -- for example, \cite{Coenders:22} show how a small set of PLRs or ALRs can be chosen in the context of generalized linear models to predict a response variable -- in the latter case of ALR selection, this leads to the choice of a subcomposition of parts.
ILR balances are not at all convenient for such supervised learning analyses since there are too many possibilities (see Supplementary Material, Section S1). 
Even if optimal balances could be chosen in cases where the number of parts is low, their interpretation in a model remains problematic in classification and regression trees, for example, where inequality conditions at the decision nodes in a tree would have to be interpreted in terms of ratios of geometric means.
If grouping of geochemical elements makes sense, then logratios of predetermined amalgamated parts would be easier to interpret and a more natural way of contrasting groups of elements in geochemistry \citep{GreenacreGrunskyBaconShone:20, Smithson:22}.
Amalgamations also alleviate the problem of data zeros, which remain a problem in geometric means.

The kimberlite data analyzed here have no zero values, and hence the logratio transformations can be applied without concern for divisions by zero or invalid logarithmic transforms.
This is not the case in general, however, and the problem of data zeros usually has to be faced, since there can be anything up to 5\% zeros in geochemical compositional data, and sometimes even higher.
 What to do about zero values remains the thorniest issue in compositional data analysis, since different strategies for replacing them with positive values can lead to different results.

Data zeros can generally be classified as:
\vspace{-0.2cm}
\begin{enumerate}
   \item structural zeros -- a value for a specific element does not exist in that geochemical environment and thus the element can be dropped or samples with structural zeros can be sequestered;
   \item missing values -- one can choose to estimate a replacement value based on nearest neighbours or some other measure of ``proximity";
   \item missing values due to being below the lower limit of detection --  here a range of imputation procedures can be used, whilst respecting the limit of detection.
\end{enumerate}
%In the case of 2 and 3, the ``type" of missing value needs to be established, and that is not always easy.

\vspace{-0.2cm}

\noindent
The problem of data zeros was treated by \cite{Sanford:93} and the whole issue of zero replacement strategies has been summarized recently by \cite{Lubbe21} and the references given therein. 

An alternative approach would be to use methods that need no zero replacement. 
For example, a transformation with good properties is to combine the chi-square standardization inherent in correspondence analysis with the Box-Cox power transformation \citep{Greenacre:21, GreenacreEtAl:23}.
 This transformation, recently called the `chiPower' transformation, although neither exactly isometric (with respect to the logratio geometry) nor exactly subcompositionally coherent, has been shown by \cite{Greenacre2023chiPower} to be very close to isometric, as well as close to coherent, close enough for all practical purposes of unsupervised analysis. 
Furthermore, the chiPower transformation is also shown to be close to coherent in statistical modeling, in the sense of giving only slightly different results when subcompositions are formed. Apart from cicumventing the zeros problem, this approach allows practitioners to interpret the compositional parts rather than their ratios in supervised analysis, greatly simplifying the task.  
Hence, it could be argued that such an alternative approach is the more satisfactory one when there are data zeros, rather than creating data replacement values for the sole purpose of using logratio methods.
As \cite{lundborg2023perturbationbased} state: ``we believe that it is generally preferable
to modify the statistical procedure to fit the data rather than vice versa".
An example of the chiPower approach is given in the Supplementary Material Section S7.

This manuscript has demonstrated a systematic and defensible workflow, called GeoCoDA, for objectively assessing geochemical data in a manner that provides enhanced insight and interpretation of processes. 
The application of simple forms of logratios in unsupervised univariate, bivariate and multivariate analysis provides insight and comprehension into the relationships of the elements and the processes that they define. Bivariate plots of the original data offer understanding into the identification of stoichiometry, and the multivariate dimension-reducing method of logratio analysis and its associated biplot identifies the most significant elemental pairs that discriminate between samples and between processes.

Geochemical data derived from materials that are comprised of minerals are governed by the principles of mineral stoichiometry. The example used in this contribution, an evaluation of the Star kimberlite lithogeochemical data, illustrates the systematic steps that can be employed to effectively evaluate and interpret the data from which geochemical processes can be discovered and subsequently validated. We have demonstrated that the five kimberlite phases (Cantuar, Pense, early-, mid- and late-Joli Fou) describe distinct geochemical properties that can be uniquely characterized and classified using standard data analytics.

The essential step in GeoCoDA, which differentiates compositional data analysis from regular statistical methodology, is the choice of the data transformation (Section 3.1).
Logratios in the forms of both the additive logratio (ALR) transform and the centered logratio (CLR) transform provide mechanisms for determining which elements influence inherent processes that exist within the data structure. The ALR transform is particularly useful for identifying processes that are governed by specific elements, or groups of elements, and is akin to the Pearce element ratio (PER) approach. 
In addition, it is important to recognize groups of compositional parts that have meaningful groupings.
In this study, we have shown that the elements Mg, Cr, Co, Ni identify processes associated with mantle contamination; the elements, Na, K, Rb, Al are associated with the process of crustal contamination; and the elements, La, Th, P, Ca, Nb, Zr, Y, Er, Yb are associated with kimberlite fractionation.
The CLR transform, equivalent to the complete set of pairwise logratios, offers the advantage of using all of the variables (elements) and assessing their relationships within and between the contained processes, and leads to the same identification of the processes described above.

The application of amalgamations further strengthens the identification of the three primary processes that are manifest in the data. Figures 4 and 5 clearly show the unique geochemical properties of the five kimberlite phases and their association with mantle/crust contamination and kimberlite fractionation. Cluster analyses, in both hierarchical and non-hierarchical forms, show statistically distinct groups that generally, but not always, coincide with the five kimberlite phases. A possible reason for the discrepancy may be due to a similarity of kimberlite phase composition (e.g. Pense/eJF or mJF/lJF) but different mineralogy, levels of contamination, or unique place in the stratigraphy of the kimberlite eruptions. The application of cluster analysis supports the choices that can be made in recognizing processes with an unsupervised approach.

A supervised multivariate approach to classification and prediction of the kimberlite phases is an important aspect of the GeoCoDA approach. In this study we have shown that the use of analysis of variance, logistic regression and classification trees reveals the distinct differences between the five phases of the Star kimberlite. Using simple pairwise ratios and logistic regression, there is a unique  distinction between the eJF and the other phases with a better than 93\% prediction accuracy based on the use of cross-validation to minimize any sampling bias. The results are shown graphically in Figures 11 and 12.

A classification tree approach provides a better prediction in terms of phase discrimination, with the advantage that the classification tree shows the critical decision points at each node of the tree. The discrimination of the eJF with the other phases and the prediction of the five phases all show high degrees of correct classification. Finally, the application of a random forest classification shows even better results. These results are nearly identical to those of \cite{Grunsky:30} where a linear discriminant analysis, using the first seven dimensions of a logratio analysis, was applied to classify the kimberlite phases. The degree of misclassification can be attributed to two possible factors, 1) the initial classification is incorrect, or 2) there is compositional overlap between the classes. The use of these four methods of prediction/classification demonstrate the unique geochemical characteristics of each of the Star kimberlite phases.

We conclude that the GeoCoDA methodology and workflow described herein provide a systematic and comprehensive way of evaluating geochemical data to enable the discovery and validation of geochemical processes.

\section*{Availability of data and R code}
Our workflow uses software tools made freely available for the \textsf{R} programming language \citep{R:21} and the package \texttt{easyCODA}. 
Data and \textsf{R} code to perform the analyses of the Star kimberlite data reported here will be available at \texttt{https://github.com/michaelgreenacre/CODAinPractice}.
%An \textsf{R} package \texttt{GeoCoDA}, which encapsulates all the methods more conveniently using high-level functions, is in progress of construction.

\section*{Acknowledgements}
BAK acknowledges support from Natural Resources Canada -- Geological Survey of Canada and the Targeted Geoscience Initiative (TGI), and Star Diamond Corp.

\newpage

\bibliography{GeoCoDA}  % PUT BACK
%%%%%%%%%%%%%%%%%%%%%%%%%%%%%%%%%%%%%%%%%%%%%%%%%%%%%%%%%%%%%%%%%%%%%%
%  SUPPLEMENTARY MATERIAL
%%%%%%%%%%%%%%%%%%%%%%%%%%%%%%%%%%%%%%%%%%%%%%%%%%%%%%%%%%%%%%%%%%%%%%
\newpage   % NEW

\centerline{\LARGE SUPPLEMENTARY MATERIAL}
\setcounter{table}{0}
\renewcommand{\thetable}{S\arabic{table}}

\setcounter{figure}{0}
\renewcommand{\thefigure}{S\arabic{figure}}

%\newpage   %%% RESTORE WHEN BACK

\bigskip

\bigskip

%%%%%%%%%%%%%%%%%%%%%%%%%%%%%%%%%%%%%%%%%%%%%%%%%%%%%%%%%%%%%%%%%%
\noindent
\textbf{\large S1. Definitions of logratios}

\medskip

\noindent
Suppose $x_j,\  j=1,\ldots,J$ is a $J$-part composition that contains no zero values: 

\centerline{$x_j >0, j=1,\ldots,J \ {\rm and} \ \sum_j x_j = 1 $.}

\medskip
\noindent
\textit{Pairwise logratio \textrm{(PLR)}}

\noindent
The simplest logratios are the pairwise logratios of two parts $j$ and $k$
\begin{equation*}
{\rm PLR}(j,j^\prime) = \log(x_{j}/x_{k})    
\end{equation*}
There are $J(J-1)/2$ unique PLRs.

\medskip
\noindent
\textit{Additive logratio \textrm{(ALR)}}

\noindent
Additive logratios are a special case of pairwise logratios where the denominator part (also called the reference part) is fixed (here it is the $J$-th part).
\begin{equation*}
{\rm ALR}(J) = \log(x_{j}/x_{J}), \quad j=1,\ldots, J-1    
\end{equation*}
There are $J$ choices for the reference part, each of which gives $J-1$ ALRs.

\medskip
\noindent
\textit{Centred logratio \textrm{(CLR)}}

\noindent
Centered logratios divide each part by the geometric mean of all the parts.
\begin{equation*}
{\rm CLR} = \log(x_{j}/(x_1 x_2 \cdots x_J)^{1/J}), \quad j=1,\ldots, J    
\end{equation*}
There are $J$ CLRs and the $j$-th one is the average of all the PLRs $\log(x_j/x_{k})$, for $k = 1, 2, \ldots, J$, one of which, $\log(x_j/x_{j})$ is zero. 

\medskip
\noindent
\textit{Isometric logratio \textrm{(ILR)}}

\noindent
Isometric logratios are ratios of geometric means of two distinct subsets of parts, with a scaling factor.
Suppose the first subset is denoted by $J_1$ and the second by $J_2$, with $|J_1|$ and $|J_2|$ numbers of parts respectively.
Then the corresponding ILR is
\begin{equation*}
\label{ILR}
 \textrm{ILR}(J_1,J_2) =  \sqrt{\frac{J_1 J_2}{J_1+J_2}} \log\frac{[\prod_{j\in J_1} x_j]^{1/J_1}}{[\prod_{j\in J_2} x_j]^{1/J_2}}
%  = \sqrt{\frac{J_1 J_2}{J_1+J_2}}\left(\frac{1}{J_1}\sum_{j\in J_1}\log(X_j)-\frac{1}{J_2}\sum_{j\in J_2}\log(X_j)   \right)
\end{equation*}
The problem is how to choose an ILR. Already for 22 parts, as in our kimberlite data, there are over two million ways of splitting the 22 parts into two disjoint subsets, without omitting any parts.
So the number of ways soars to billions from which to choose a single ILR with two disjoint sets with reduced numbers of parts, of which all the PLRs, ALRs and even CLRs are special cases, apart from the scalar multipliers.
The way ILRs are usually defined is by constructing a dendrogram based on inter-part distances, which is very cheap since this is done using an agglomerative clustering algorithm. 
This solution results in $J-1$ ILRs, called the ``default" ILR, but there is no reason why this particular choice should be related to the research objective, especially in the context of supervised learning. 

\medskip
\noindent
\textit{Summed logratio \textrm{(SLR)}(or amalgamation logratio) }

\noindent
Amalgamation logratios are ratios of sums of two distinct subsets of parts, and require no scaling factor.
Again, suppose the first subset is denoted by $J_1$ and the second by $J_2$, with $|J_1|$ and $|J_2|$ numbers of parts respectively.
Then the corresponding amalgamation logratio is
\begin{equation*}
\label{SLR}
 \textrm{SLR}(J_1,J_2) =  \log\frac{\sum_{j\in J_1} x_j}{\sum_{j\in J_2} x_j}
\end{equation*}
SLRs are much easier to understand and interpret than ILRs, since they are regular PLRs involving sums of parts. 
They can truly be called ``balances" in the intuitive sense of aggregated parts being compared with other aggregated parts. 
SLRs are usually computed using knowledge-driven combinations of parts, as was done in the definition of mantle, crustal and kimberlite amalgamations in Section 3.2.
Similarly, in fatty acid analyses in biochemistry, the fatty acids are typically amalgamated in three subsets of saturated, monounsaturated and polyunsaturated fatty acids, and ratios are computed between these amalgamations.

% \newpage
\bigskip

\bigskip

\noindent
\textbf{\large S2. Compositional barplot of kimberlite data, original untransformed percentages}

\begin{figure*}[htbp]
\begin{center}
\includegraphics[width=13cm]{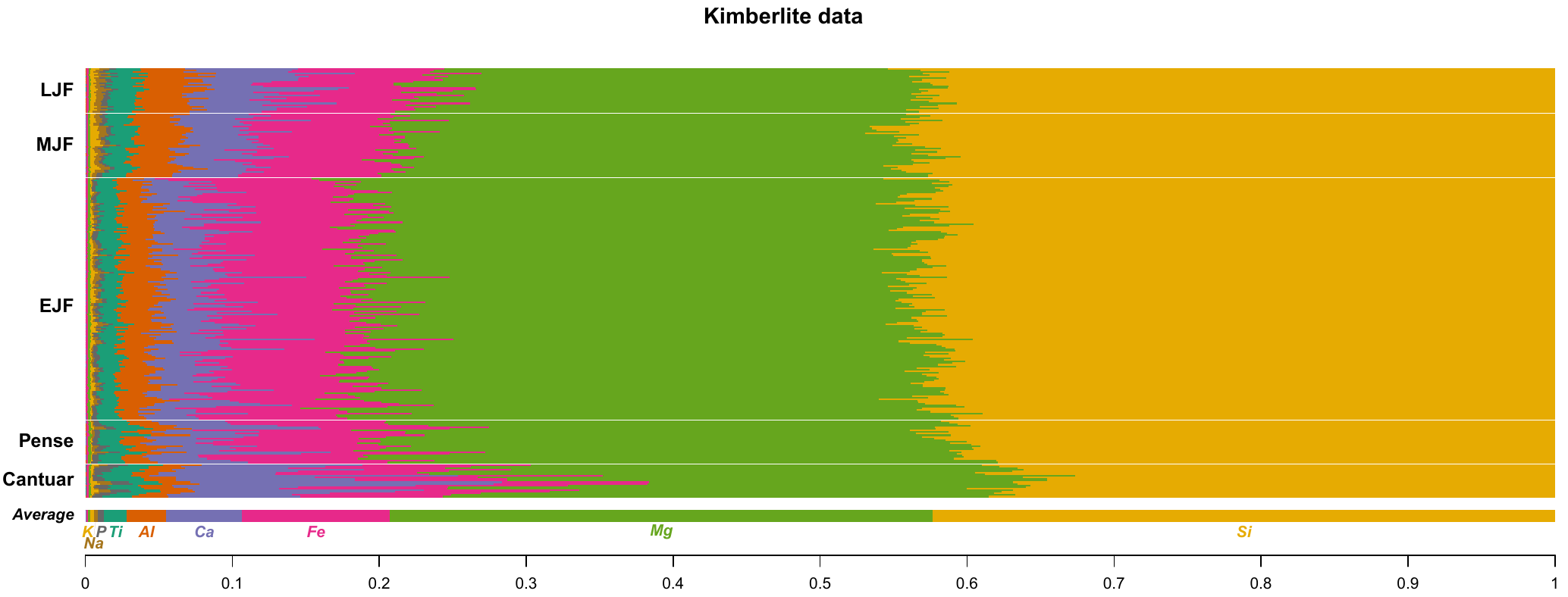}
\caption{An attempt to show all the compositional data, using a linear scale, in the form of a compositional bar plot, where elements have been ordered in increasing mean value. Only the most common major elements can be seen clearly, whereas all other elements with much smaller values are difficult, or impossible, to distinguish. The five kimberlite phases have been ordered from oldest in the bottom rows to youngest at the top. Figure 2 of the paper shows all elements using a nonlinear scale, where the rarer elements have an amplified scale and the abundant elements a shrunken scale.
%The nonlinear transformation in Figure 2 was achieved by (i) multiplying the compositional data by the inverse of the smallest value in the whole data set, which makes the smallest value 1 while preserving all the relative values (i.e., ratios) i the data, then (ii) performing a logarithmic transformation to pull down all the high values, and finally (iii) closing the data set.
}
\label{BarPlot}
\end{center}
\end{figure*}

\newpage
\noindent
\textbf{\large S3. Additional possible amalgamations}

\begin{figure*}[htbp]
\begin{center}
\includegraphics[width=13cm]{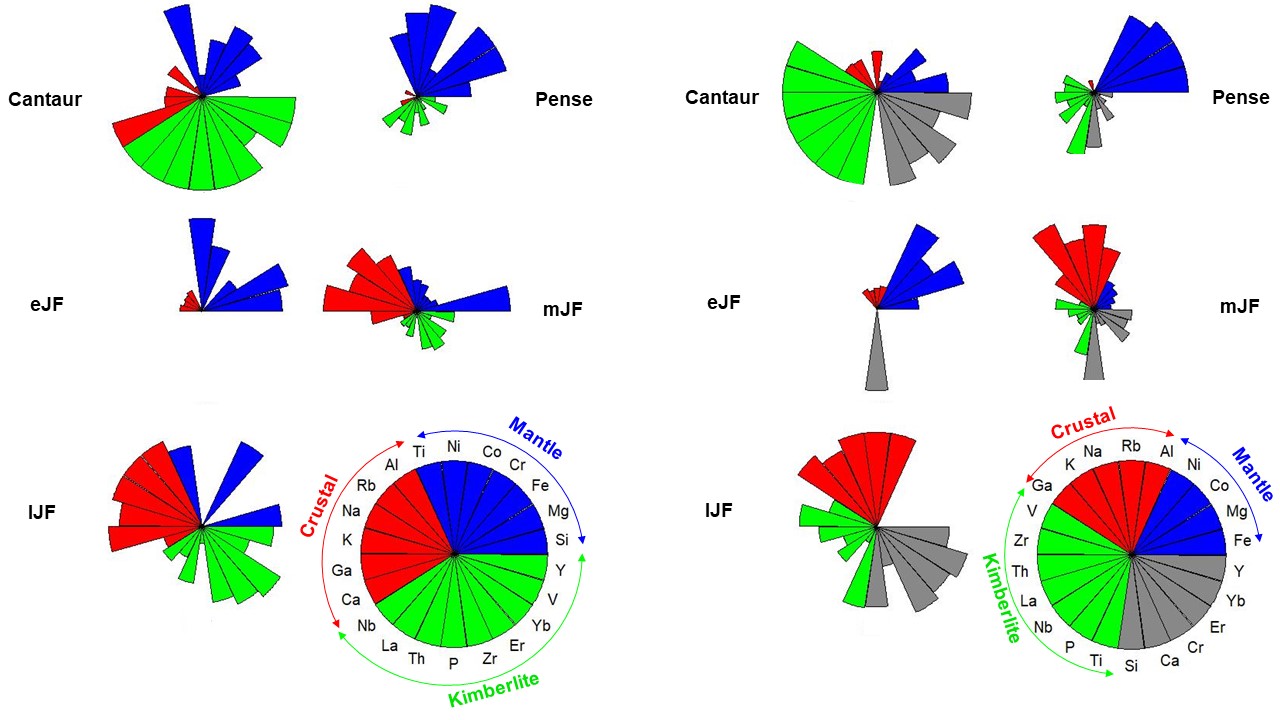}
\caption{Two additional attempts at defining amalgamations based on domain knowledge. These are less successful in separating the kimberlite phases than the one used in Section 3.3, Figure 4. }
\label{AdditionalStarPlots}
\end{center}
\end{figure*}

\newpage
\noindent
\textbf{\large S4. Grid lines in the centered ternary diagram}
\bigskip
\begin{figure*}[htbp]
\begin{center}
\includegraphics[width=12.1cm]{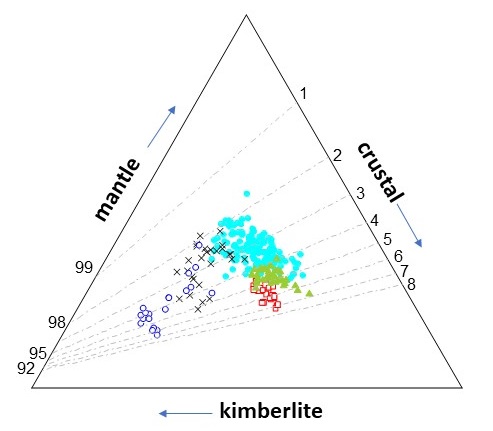}
\caption{Centered ternary plot of Figure 5B, showing one set of grid lines, which are still straight but the scales on the side have been deformed in a nonlinear monotonic way. The lines show how the scale of mantle amalmgamation has been compressed and the scale of crustal amalgamation expanded to spread out the sample points.}
\label{CenteredTernary}
\end{center}
\end{figure*}

\newpage
\noindent
\textbf{\large S5. Alternative clusterings of the samples}

\medskip
\noindent
The table below shows the results of hierarchical clustering of the 270 kimberlite samples, using the ALR transformation with Zr as the fixed denominator element, followed by Ward clustering.
The cross-tabulation of the five-cluster solution with the five kimberlite phase groups is given in Table \ref{CrossTable5}. 

\begin{table}[h]
\begin{tabular}{crrrrr}
Cluster &  Cantuar & Pense & \quad eJF & \quad mJF & \quad lJF \\
\hline
  1   &     0  &    0 &  66 &   1 &   0 \\
  2   &    13  &   10 &  12 &   0 &   0 \\
  3   &     2  &   17 &  40 &   2 &   0 \\
  4   &     0  &    0 &  33 &   0 &   0 \\
  5   &     6 &     0 &   3 &  37 &  28 \\
\hline
\end{tabular}
\caption{Cross-tabulation of five clusters of the samples, obtained by hierarchical clustering on the ALR-transformed data with reference Zr, versus their phase classes.}
\label{CrossTable5}
\end{table}
\noindent
Based on assigning the clusters to the most frequent phase group, the assignment is 70.0\% successful (189 out of 270 assignments, but no assignment to Pense or lJF groups).

Table \ref{CrossTable6} shows the results of hierarchical clustering of the same samples, using the chi-square distances in correspondence analysis, and cross-tabulating the five-cluster solution with the five kimberlite phase groups. 

\begin{figure*}[h]
\begin{center}
\includegraphics[width=12.1cm]{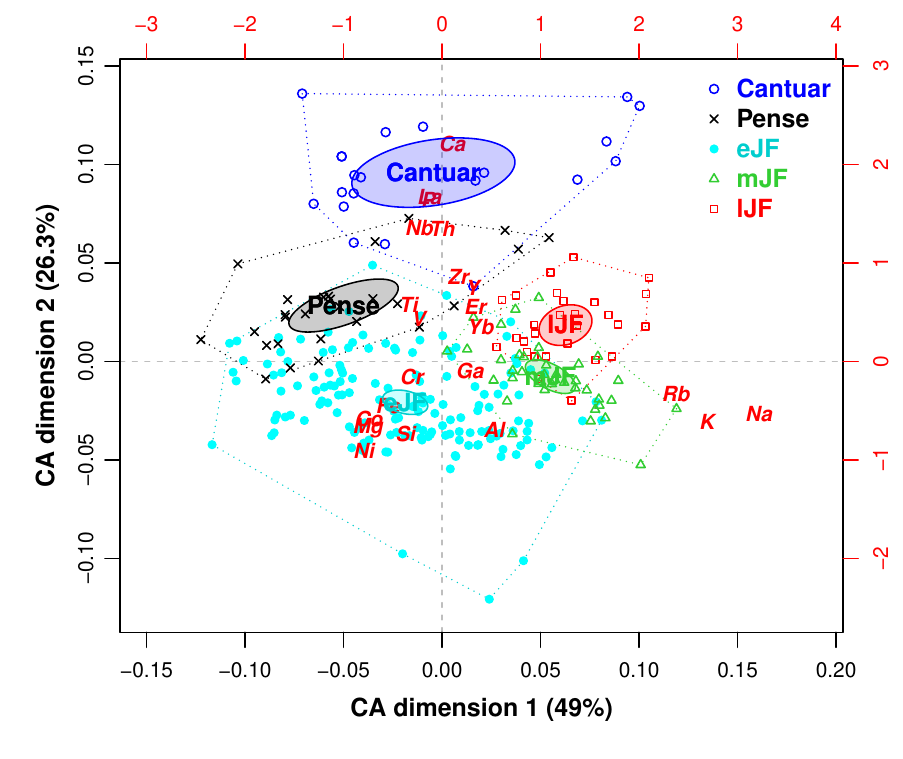}
\caption{Correspondence analysis of the kimberlite data, fourth-root transformed. Variance (i.e., inertia) explained in this two-dimensional solution is 75.3\%. This result is very similar to that of the logratio (LRA) in Figure 9, also that of the PCA of the ALRs with respect to Zr in Figure 10. The Procrustes correlation of this CA geometry with that of the logratio geometry is 0.972 and would increase with a stronger power transformation, tending to 1 as the power tends to zero, since the data are all strictly positive.}
\label{BiplotCA}
\end{center}
\end{figure*}

\begin{table}[h]
\begin{tabular}{crrrrr}
Cluster &  Cantuar & Pense & \quad eJF & \quad mJF & \quad lJF \\
\hline
  1   &     0  &  19  & 125 &    5 &   0 \\
  2   &     2  &   7 &   17 &    6 &  14 \\
  3   &     0  &   0 &    9 &   29 &  14 \\
  4   &    13  &   1 &    3 &    0 &   0 \\
  5   &     6 &     0 &   0 &    0 &   0 \\
\hline
\end{tabular}
\caption{Cross-tabulation of five clusters of the samples, obtained by hierarchical clustering on the chi-square distances, versus their phase classes.}
\label{CrossTable6}
\end{table}
\noindent
Based on assigning the clusters to the most frequent phase group, the assignment is 70.4\% successful (190 out of 270 assignments, but again no assignment to Pense or lJF groups).
If the fourth-root transform is applied to the compositions, as described in Section S7 (see Figure \ref{BiplotCA}), then the percentage rises to 73.0\% (197 out of 270), again with no assignments to Pense and lJF.

%to be incorporated
 In summary, for the ALR distances there are 70.0\% correct assignments, and for chi-square distances
%: $V = 0.580$, with 
70.4\% (without power transform) and 74.0\% (with power transform) correct assignments .
These three clusterings all improve on the regular logratio distance clustering, and they also assign kimberlite phase eJF more correctly.

\newpage
\medskip
\noindent
\textbf{\large S6. Procrustes correlations for all ALR transformations}
\smallskip
\begin{table*}[ht]
\begin{center}
\hskip0.5cm\begin{tabular}{rcc}
\  & \  & Procrustes \\[-2pt]
Order & Element & correlation \\
\hline
 1 &  Zr &  0.977 \\ 
 2 &  Y  &  0.977 \\ 
 3 &  Cr &  0.973 \\ 
 4 &  V  &  0.972 \\ 
 5 &  Fe &  0.968 \\ 
 6 &  Si &  0.966 \\ 
 7 &  Nb &  0.962 \\ 
 8 &  Er &  0.961 \\ 
 9 &  Ti &  0.959 \\ 
10 &  Mg &  0.955 \\ 
11 &  Ga &  0.955 \\ 
12 &  Co &  0.945 \\ 
13 &  Th &  0.944 \\ 
14 &  La &  0.941 \\ 
15 &  Yb &  0.938 \\ 
16 &  Al &  0.937 \\ 
17 &  P  &  0.919 \\ 
18 &  Ni &  0.918 \\ 
19 &  Rb &  0.889 \\ 
20 &  K  &  0.877 \\ 
21 &  Na &  0.835 \\ 
22 &  Ca &  0.823 \\
\hline\\
\end{tabular}
\caption{List of elements used as references in ALR transformations. The elements are in descending order of Procrustes correlation with the exact logratio geometry. Based on these results, element Zr was chosen for the ALR transformation in Figure 10.}
\label{ALR_Procrust}
\end{center}
\end{table*}

\newpage
\noindent
\textbf{\large S7. Correspondence analysis of power-transformed compositions}

\bigskip
There is a long history of correspondence analysis being used to analyse geochemical data -- see, for example, \cite{David:73,David:77, Grunsky:86}.
CA weights the elements proportionally to their mean values, so that 
the major elements will play a stronger role in the analysis.
This means that CA is actually analogous to weighted logratio analysis, where the CLRs are weighted by these respective means \citep{Greenacre:10a, GreenacreLewi:09}.
The contribution of each element to the total weighted variance in CA, called inertia, can be measured: it turns out that, in our earlier terminology, element Ca contains 59.0\% of the total inertia, and would thus dominate the CA, which is an analysis designed to be inertia-explaining. 

The interesting feature of CA is that for strictly positive data, a CA of power-transformed data tends to the (unweighted) LRA solution as the power $\lambda$ tends to zero \citep{Greenacre:10a}.
Power transformation changes the relative contributions of the elements to the total inertia by making the mean element values more uniform.
With a fourth-root transform the contribution of element Ca to the total inertia is reduced from 59.0\% to 15.2\% of the total, and the CA solution shown in Figure S4 is quite similar to the ones in Figures \ref{BiplotLRA} and \ref{BiplotPCA}.
The advantage of CA is that it can be applied to data with zeros, but there the choice of the power $\lambda$ has to be done to maximize the subcompositional coherence -- see \cite{Greenacre:11a,GreenacreEtAl:23}.
In the present case, where all the data are strictly positive, subcompositional coherence is achieved as $\lambda$ tends to zero, where CA tends to LRA. 
An optimality criterion for determining a $\lambda$ value that is not equal to 0 is only possible if some external condition is imposed: for example, optimizing the separation of the phases in the two-dimensional result.
It turns out that $\lambda=0.88$, which is a mild power transformation, optimally separates the phase groups. 
For the regular CA, the between-group sum-of-squares is 55.5\% of the total, whereas for the power-transformed CA with $\lambda=0.88$, this percentage increases slightly to 55.8\%.
This is a small benefit for the present data where the regular CA solution already separates the phases very well.

To compare this with the best separation possible in a CA, a discriminant form of CA where the group centroids are optimally displayed in a two-dimensional projection, this percentage rises to the value of 57.1\%. 
This CA of the phase centroids is not reported since it is almost indistinguishable from the CA of individual samples.
In other applications, however, where the group separation is not so clear in the CA solution of individual samples, the analysis of group centroids rather than individual samples can show a dramatic improvement in the group separation.

The use of power transformation combined with the chi-square normalization has been called the chiPower transformation, and shown to have excellent properties for analyzing compositional data with zeros \citep{Greenacre2023chiPower}.
The advantages are that no zero replacement is needed, and the transformed variables can be interpreted in terms of the parts themselves, not their ratios.

\clearpage
\newpage
\noindent
\textbf{\large S8. Multinomial logistic model coefficients}
\smallskip
\begin{table*}[ht]
\begin{center}
\hskip0.5cm\begin{tabular}{lrrrr}
Phase & Intercept & Si/Zr \qquad & P/Rb \qquad & Si/Ni \qquad \\[-2pt]
\hline
Pense & $-$108.36 & {\bf 23.22} (6.56) & {\bf 4.96} (1.71) & {\bf $-$21.05} (7.26) \\
eJF & $-$270.96 & {\bf 38.87} (6.97) & $-$0.26 (1.68) & $-$-9.54 (6.89) \\
mJF & $-$132.33 & {\bf 14.99} (4.40) & {\bf $-$4.33} (1.62) & 5.99 (5.84) \\
lJF & $-$55.30 & 3.73 (3.74) & {\bf $-$4.21} (1.67) & 8.74 (5.97) \\
\hline\\
\end{tabular}
\caption{Results of multinomial logistic regression, predicting the five phase categories. The earliest phase, Cantuar, is used as the reference phase, so all coefficients represent differences with respect to Cantuar. Standard errors are given in parenthese. Significant regression coefficients of the logratios, according to the plus/minus two standard errors criterion, are shown in boldface.}
\label{MultinomialLogistic}
\end{center}
\end{table*}

%\bibliographystyle{sn-basic}
%\bibliographystyle{natbib}
% \bibliography{GeoCoDA}  RESTORE
\end{document}